\documentclass[12pt,preprint]{aastex}
\def\ltsima{$\;\buildrel < \over \sim \;$}
\def\simlt{\lower.5ex \hbox{\ltsima}}
\def\gtsima{$\;\buildrel > \over \sim \;$}
\def\simgt{\lower.5ex \hbox{\gtsima}}

\newcommand{\water}{H$_2$O}

\newcommand{\um}{$\:\mu$m}

\newcommand{\cmc}{cm$^{-3}$}

\newcommand{\etal}{{\em et$\:$al.}}

\shorttitle{Spitzer observations of HH54 and HH7--11}
\shortauthors{Neufeld et al.}

\begin{document}

\title{Spitzer observations of HH54 and HH7--11: mapping the H$_2$ ortho-to-para ratio
in shocked molecular gas}
\author{David A. Neufeld\altaffilmark{1}, Gary~J.~Melnick\altaffilmark{2}, 
Paule~Sonnentrucker\altaffilmark{1}, Edwin~A.~Bergin\altaffilmark{3}, 
Joel~D.~Green\altaffilmark{4}, Kyoung~Hee~Kim\altaffilmark{4}, 
Dan~M.~Watson\altaffilmark{4}, 
William~J.~Forrest\altaffilmark{4}, 
and Judith~L.~Pipher\altaffilmark{4}}

\altaffiltext{1}{Department of Physics and Astronomy, Johns Hopkins University,
3400 North Charles Street, Baltimore, MD 21218}
\altaffiltext{2}{Harvard-Smithsonian Center for Astrophysics, 60 Garden Street, 
Cambridge, MA 02138}
\altaffiltext{3}{Department of Astronomy, University of Michigan, 825 Dennison Building, Ann Arbor, MI 48109}
\altaffiltext{4}{Department of Physics and Astronomy, University of Rochester, Rochester, NY 14627}

\begin{abstract}

We report the results of spectroscopic mapping observations carried out toward the Herbig-Haro objects HH7--11 and HH54 over the $5.2 - 37\rm\,\mu m$ region using the Infrared Spectrograph of the {\it Spitzer Space Telescope}.  These observations have led to the detection and mapping of the S(0) -- S(7) pure rotational lines of molecular hydrogen, together with emissions in fine structure transitions of Ne$^+$, Si$^+$, S, and Fe$^+$.  The H$_2$ rotational emissions indicate the presence of warm gas with a mixture of temperatures in the range $400 - 1200$~K -- consistent with the expected temperature behind nondissociative shocks of velocity $\rm \sim 10 - 20\, km\, s^{-1}$ -- while the fine structure emissions originate in faster shocks of velocity $\sim 35 - 90\, \rm km \, s^{-1}$ that are dissociative and ionizing.  Maps of the H$_2$ line ratios reveal little spatial variation in the typical admixture of gas temperatures in the mapped regions, but show that the H$_2$ ortho-to-para ratio is quite variable, typically falling substantially below the equilibrium value of 3 attained at the measured gas temperatures.  The non-equilibrium ortho-to-para ratios are characteristic of temperatures as low as $\sim 50$~K, and are a remnant of an earlier epoch, before the gas temperature was elevated by the passage of a shock.  Correlations between the gas temperature and H$_2$ ortho-to-para ratio show that ortho-to-para ratios $< 0.8$ are attained only at gas temperatures below $\sim 900$~K; this behavior is consistent with theoretical models in which the conversion of para- to ortho-H$_2$ behind the shock is driven by reactive collisions with atomic hydrogen, a process which possesses a substantial activation energy barrier ($E_A/k \sim 4000$~K) and is therefore very inefficient at low temperature.  The lowest observed ortho-to-para ratios of only $\sim 0.25$ suggest that the shocks in HH54 and HH7 are propagating into cold clouds of temperature $\simlt 50$~K in which the H$_2$ ortho-to-para ratio is close to equilibrium.

\end{abstract}


\section{Introduction}

Stars have a profound effect upon the interstellar medium (ISM) that surrounds them.  Stellar radiation heats -- and sometimes photoionizes and photodissociates -- the interstellar gas, while supersonic outflows associated with stellar birth (in protostellar outflows) and death (in supernova remnants) send shock waves propagating through the ISM.  In regions of high mass star-formation -- such as the Orion Molecular Cloud, for example -- many of these processes operate simultaneously, making their separate effects hard to disentangle and elucidate.  In Herbig-Haro objects and supernova remnants, by contrast, the effects of shock waves can be studied in relative isolation in an environment where the stellar radiation field is not greatly enhanced.

Over more than three decades, the physics and chemistry of interstellar shock waves have been the subject of intensive study, both theoretical and observational, the results of which have been summarized in several review articles (e.g. Walmsley \& Pineau Des For{\^e}ts 2005; Draine \& McKee 1993).  When shock waves propagate within molecular clouds, the shock velocity and magnetic field determine the fate of the gas.  Fast shocks (with propagation velocities $v_s \simgt 40 \,\rm km \, s^{-1}$ given interstellar magnetic fields of typical strength) are dissociative: they destroy molecules and may ionize the resultant atoms.  Slow shocks ($v_s \simlt 40 \,\rm km \, s^{-1}$), by contrast, are non-dissociative: they heat and compress the molecular gas through which they pass, and may modify its chemical composition, but molecules survive their passage.  The chemical changes wrought by non-dissociative shocks typically result from elevated gas temperatures or the phenomenon of ion-neutral drift (crucially important in non-dissociative shocks), which can drive chemical processes that are negligibly slow in cold clouds because they are endothermic or possess activation energy barriers.  The result of these processes depends critically upon the timescale for key chemical reactions relative to the period of time for which shock elevates the temperature and ion-neutral drift velocities. In a previous study of HH54 carried out with the {\it Infrared Space Observatory (ISO)}, Neufeld et al.\ (1998; hereafter NMH98) argued that the H$_2$ ortho-to-para ratio, observed to lie significantly below its equilibrium value ($\sim 3$) in the warm ($T \sim 650\,$K) shocked gas in HH54, provided a valuable probe of the relevant timescales.

The Infrared Spectrograph (IRS) on board the {\it Spitzer Space Telescope} is a powerful tool for the study of shocked interstellar gas.  With a wavelength coverage of 5.2 -- 37 $\mu$m, {\it Spitzer}/IRS provides access to the S(0) through S(7) pure rotational lines of H$_2$, as well as fine structure emissions of several atoms and ions produced in fast, dissociative shocks: Ne$^+$, Si$^+$, S, and Fe$^+$.  The great sensitivity of {\it Spitzer}/IRS, together with the multiplex advantage inherent in long slit spectroscopy, allows large regions to be mapped rapidly.  

In this paper we report the results of spectral line mapping in HH54 and HH$7-11$, two well-studied Herbig-Haro objects known to contain warm molecular gas.  HH 7--11 is a chain of Herbig-Haro objects in the Perseus molecular cloud, associated with NGC 1333 (Snell \& Edwards 1981).  These objects lie in the blue-shifted part of a CO outflow that originates near the protostar SVS 13.  HH 7--11 has been studied extensively through observations of optical emission lines (Hartigan, Curiel, \& Raymond 1989), near-IR emission lines of [Fe~II] (Stapelfeldt et al 1991; Gredel 1996), H$_2$ vibrational emissions (Noriega-Crespo et al. 2002, Khanzadyan et al.\ 2003), far-infrared atomic and molecular lines (Molinari et al. 2000); and through high resolution studies of CO emission (Bachiller et al. 2000).   One key result to emerge from previous studies is that the hot H$_2$ emission exhibits a classic bow shaped morphology in HH~7, with an emission peak offset from the peak emission of atomic fine structure lines.   Observations of the HH7--11 flow have proven valuable in constraining detailed models for shock excited gas  (e.g. Hartigan, Raymond, \& Hartmann 1987; Smith, Khanzadyan, \& Davis 2003).  The distance to HH7-11 is a matter of debate; we adopt the value of 250 pc used by Enoch et al. (2006; see discussion in that paper which refers to \u{C}ernis 1993; \u{C}ernis \& Strai\u{z}ys 2003; Belikov et al. 2002). HH54 is located in the Chamaeleon II molecular cloud at an assumed distance of 200 pc (Hughes \& Hartigan 1992), and is known to have a large column of warm ($>$ 2000 K) shocked molecular and atomic gas (Schwarz \& Dopita 1980; Gredel 1994).  While it is less well studied than HH7-11, HH54 is notable because of an unusually large abundance of water vapor (Liseau et al. 1996) and H$_2$ ortho-to-para ratios that are below the equilibrium value expected at the observed gas temperature.

Our observations and data reduction methods are discussed in \S 2 and Appendix A, and the results are presented in \S 3.  The implications of these results for our understanding of interstellar shock waves are discussed in \S 4.  A brief summary follows in $\S 5$.

\section{Observations and data reduction}

One field in HH54, and two fields in HH7--11 -- one centered on HH7 and one on the protostar SVS13 -- were observed using the IRS as part of the IRAC Guaranteed Time program.
The Short-Low (SL), Short-High (SH) and Long-High (LH) modules were used to provide complete spectral coverage over the wavelength range available to IRS.  Maps covering fields of size $\sim 1^\prime \times 1^\prime$ were obtained by stepping the slit perpendicular, and -- in the case of SH and LH -- parallel to its long axis.  The observational details are provided in Table 1.

The data were processed at the Spitzer Science Center (SSC), using version 12 of the processing pipeline, and then reduced further using the SMART software package (Higdon et al.\ 2004), supplemented by additional routines that we have developed.
The key capabilities provided by these additional routines were the removal of bad pixels in the LH and SH data, the calibration of fluxes obtained for extended sources,
the extraction of individual spectra at each position sampled by the IRS slit, and the creation of spectral line maps from the set of extracted spectra.   
Details of our data reduction method, as well as the various tests we performed to validate the additional routines that we developed, are discussed in Appendix A. 

\section{Results}

\subsection{Spectral line maps}

Widespread rotational emissions from molecular hydrogen, along with fine structure emissions from atomic sulfur and
several singly-ionized species, 
were detected in each of the two sources we mapped.  
In Figures 1 and 2 we present maps of the H$_2$ line emissions detected in HH54.  The straight lines demark 
the regions within which each transition was mapped. The S(1) through S(7) lines all show morphological similarities, with most maps showing local maxima close to clumps E and K and to clump C (in the designation of Sandell et al.\ 1987).  However, real differences in the spatial distribution of the various transitions are apparent and provide key information about the spatial variation of the H$_2$ excitation and ortho-to-para ratio; these variations are discussed in detail 
in \S 4.2 and \S 4.4 below.

Emissions in several fine structure transitions have also been detected toward HH54.  Figure 3 shows the distribution of the four fine structure emission lines
that are strong enough to map:  [NeII] 12.8$\mu$m, [FeII] 26$\mu$m, [SI] 25$\mu$m, and [SiII] 35$\mu$m. 
Offsets between the fine structure emissions and the H$_2$ emissions are clearly evident in Figure 7 (left panel), which shows contour plots that compare the distributions of the [NeII] and H$_2$ S(3) emissions, suggesting a different origin for the H$_2$ and fine structure emissions, discussed further in \S 4.1 

Figures 4 and 5 show maps of the H$_2$ line emissions detected in HH7--11.  As before, the straight lines demark 
the regions within which each transition was mapped; in this case, two overlapping rectangular fields were mapped
in the SL and SH modules and one field in the LH module.  The dashed circles centered at offset (0,0) encircle the regions 
close to the bright continuum source SVS13 where the line intensity estimates are unreliable.  The S(1), S(2), 
S(3) and S(5) line maps are remarkably extended, with diffuse emission covering (and extending up to 20$^{\prime \prime}$ on either side of) the line joining HH7 to SVS13.  A detailed  discussion of the differences in the spatial distribution of the various transitions -- and of the inferred variations in the H$_2$ excitation and ortho-to-para ratio -- is again deferred to \S 4.2 and \S 4.4.

As in HH54, emissions in several fine structure transitions have also been detected in HH7--11.  Figure 6 shows maps of the 
[FeII]~26$\mu$m, [SI] 25$\mu$m, and [SiII] 35$\mu$m line emissions detected with the LH module toward HH7.  Unlike in HH54, the [NeII] 12.8~$\mu$m line is not detected.  Figure 7 (right panel) shows that the fine structure emissions, as traced by [SiII] in this case, are again offset from the H$_2$ emissions, peaking several arcseconds closer to SVS13.

\subsection{Average spectra}

We have averaged the observed spectra over several regions within the HH54 and HH7--11 fields, thereby obtaining substantial improvements in the signal-to-noise ratio.  The resultant spectra, plotted in Figure 8, are the averages of all spectra within a series of circular regions, the contributing spectra being weighted by a Gaussian taper (HPBW of 15$^{\prime\prime}$) from the center of each synthesized beam.  The position of each synthesized beam is demarked by the dotted circles in Figure 7, and the central positions are given in the caption to Figure 8. 

In HH54, we obtained average spectra for circular regions centered on the peak of the [NeII] emission and two peaks of the H$_2$ emission.  We denote the three circular beams HH54FS (the peak of the [NeII] emission), HH54C (the northern of the two H$_2$ peaks, being roughly coincident with clump C identified by Sandell et al.\ 1987), and HH54E+K (the southern H$_2$ peak, roughly coincident with clumps E and K).
In HH7--11, we obtained the average spectrum for a circular region centered on the HH7 bow shock. 

The spectra plotted in Figure 8 all show very weak continua, leading to large line-to-continuum ratios even in the low-resolution module (Short-Low, left panel, with $\lambda/\Delta \lambda$ of only $\sim\,$60).  A fairly rich spectrum of H$_2$ rotational lines and low-excitation fine structure lines is apparent in each case.

A total of eight H$_2$ rotational lines and eight fine structure lines of Ne$^+$, Si$^+$, S, and Fe$^+$ (5 transitions) have been detected at one or more such region.  Table 2 lists the detected lines together with the mean frequency-integrated intensities for each circular region.  As discussed in \S 4.1 below, the H$_2$ line ratios and intensities are typical of slow nondissociative shocks, while the fine structure line intensities are in good agreement with those expected from fast dissociative shocks.

No fewer than five [FeII] fine structure lines are clearly detected toward position HH54FS.  The observed spectra are shown in Figure 9, and the transitions detected are marked on the Grotrian diagram in Figure 10.  The strength of the higher excitation transitions originating in the $^4F$ term, relative to the lower transitions within the $^6D$ term, is expected to be an increasing function of both temperature and density.  As discussed in \S 4.1, the [FeII] line ratios provide a valuable probe of the physical conditions in the emitting region. 

Upper limits are also obtained for two rotational transitions of water vapor that lie within the LH band and are discussed in \S 4.3 below.
In addition, the R(3) and R(4) rotational lines of HD have been tentatively detected and 
discussed elsewhere (Neufeld et al.\ 2006).

\section{Discussion}

\subsection{Fast and slow shocks}

In both sources that we observed, clear offsets between the fine-structure emissions and the molecular hydrogen emissions are evident.  Similar offsets have also been observed toward Herbig-Haro objects in the GGD37 (Cepheus A West) region  (Raines, 2000).  The fine structure emissions in HH7 are offset by $\sim 5^{\prime\prime}$ from the bow shock, corresponding to a projected distance of $\sim 2 \times 10^{16} \rm \, cm$ for an assumed distance to the source of 250~pc.
Such an offset has been observed previously in HH7 (Staplefeldt et al. 1991), where near-infrared emissions from [FeII] and H$_2$ have been mapped at subarcsecond resolution by Noriega-Crespo et al.\ (2002) using the {\it Hubble Space Telescope}.  Noriega-Crespo et al.\ interpreted the [FeII] emissions as emerging from a ``Mach disk'', in which the collimated supersonic outflow is decelerated and heated in a fast shock.  In this picture, the fast shock is offset from a slower non-dissociative bow shock that is driven into the ambient molecular medium, the latter giving rise to the molecular hydrogen emissions observed from HH7.  Our mid-infrared observations indicate that the [SiII] and [SI] fine structure emissions -- like the [FeII] emissions -- also emerge from a region offset from the bow shock.  The morphology of the [SiII] map is most similar to the [FeII] map.  These observations support the two-shock picture in which the [FeII], [SI] and [SiII] fine structure emissions emerge from fast shock.  

Our observations of HH54 indicate a similar offset ($\sim 10^{\prime\prime}$, corresponding to a projected distance of $\sim 3 \times 10^{16} \rm \, cm$ for a source distance of 200~pc) between the fine structure and molecular hydrogen emissions.  As in HH7, the centroid of the fine structure emissions lies closer to the presumed source of the outflow than the H$_2$ emissions, consistent with the two-shock picture in which the fine structure emissions trace a fast shock that decelerates the collimated outflow.

We have compared the observed fine structure line intensities listed in Table 2 with predictions of the fast shock model of Hollenbach \& McKee (1989, hereafter HM89).  Because the
beam filling factor of shocked material is not known, line ratios provide a more valuable diagnostic than absolute intensities.   Although the predictions of the HM89 shock models are undoubtedly somewhat uncertain -- given substantial uncertainties in the collisional excitation rates for the various transitions and in the gas-phase abundances of the various elements -- the predicted [NeII]~12.8$\,\mu$m/[FeII]~26.0~$\,\mu$m line ratio is a strongly increasing function of the shock velocity, values $\simgt \rm 60\,km\,s^{-1}$ being needed to ionize Ne significantly.  The [NeII]~12.8$\,\mu$m/[FeII]~26.0$\,\mu$m line ratio of $\sim 0.6$ observed in HH54 requires a shock velocity $\sim \rm 80 \, km \,s^{-1}$.  In HH7, by contrast, the absence of detectable [NeII]~12.8$\,\mu$m emission places an upper limit $v_s < \rm 70\,km\,s^{-1}$, whilst our measurement of the [FeII]~17.9$\,\mu$m transition requires a shock velocity\footnote{Considerably smaller shock velocities would be obtained in a comparison with the model predictions of Hartigan et al.\ (1989), which apply to shocks in low density clouds that are assumed to be ionized prior to the passage of the shock.} of at least $35\,\rm km\,s^{-1}$.

The HM89 shock models predict that the [FeII] and [SiII] line intensities show a very similar dependence upon the shock velocity, whilst the [SI]/[FeII] and [SI]/[SiII] line ratios are a decreasing function of shock velocity.
This prediction provides a natural explanation for the fact that the [SiII] map is most similar to the [FeII] map: insofar as the fine structure emissions may arise from a superposition of shocks at varying velocities, the [SI] emissions would be expected to show a spatial distribution that is somewhat different from [SiII] and [FeII]. 

An independent check of the two-shock picture is provided by our detection of multiple [FeII] lines. 
The mimimum temperature derived from the [FeII] line ratios ($\sim 5000$~K) is larger than the maximum temperature at which H$_2$ can survive for any appreciable time period -- and is a factor $\sim 5$ larger than that inferred from the H$_2$ line ratios (see \S 4.2 below) -- lending strong support to the theory that both dissociative and nondissociative shocks are present in the source.

\subsection{The temperature and column density of shocked molecular hydrogen}

The H$_2$ S(0) -- S(7) emissions provide a valuable probe of the temperature distribution in the shocked gas in which H$_2$ is the dominant constituent.  The observed line intensities for these H$_2$ quadrupole transitions provide a direct measurement of the column density, $N(J_u)$ in each upper state, $J_u$, because all the lines involved are optically-thin.  As in NMH98, we applied extinction corrections based upon the interstellar extinction curves of Weingartner \& Draine (2001; ``Milky Way, $R_V=3.1$").  For HH54, we adopted Gredel's (1994) estimate of 0.3 for $E(J-K)$, obtained from observations of the [FeII] 1.257 and 1.644 $\mu$m transitions toward HH54E+K, whilst for HH7 we adopted the estimate $E(J-K) = 0.7$ obtained by Gredel (1996) for the nearby source HH8.

In Figure 11, based upon the line intensities tabulated in Table 2, we present rotational diagrams for each of the four circular regions for which we computed an average spectrum (\S 3.2).  Two features are common to all observed positions.  First, as discussed further in \S 4.4 below, the rotational diagrams all show the ``zigzag" patterns characteristic of non-equilibrium ortho-to-para ratios, in which a fit to the para- (even-$J$) states lies above that for the ortho- (odd-$J$ states).  Second, the rotational diagrams all exhibit a degree of curvature, with the slope of the rotational diagram decreasing with increasing $J$.  This curvature indicates that each beam is sampling material at a range of temperatures.  Furthermore, the absence of any turnover in the curvature at high $J$ suggests that the rotational level populations are close to local thermodynamic equilibrium (LTE) even for $J$ as high as 9.

For each circular region in Table 2, we determined an excitation temperature, $T_{XY}$, from the ratio of the $J=X$ and $J=Y$ column densities and defined according to
$$T_{XY} = {E_{Y} - E_{X}  \over k\, {\rm ln}\,[N_{X} g_{Y}/ N_{Y} g_{X}]}$$   
where $E_J$ and $g_J$ are the energy and degeneracy of state $J$. Because the ortho-para ratio shows departures from equilibrium, the quantity 
$T_{XY}$ is most useful when $X$ and $Y$ are either both odd or both even.  
The excitation temperatures $T_{42}$, $T_{53}$, $T_{64}$, $T_{75}$, $T_{86}$ and $T_{97}$ are given in Table 3;  the inequalities $T_{42} < T_{64} < T_{86}$ and
$T_{53} < T_{75}  < T_{97}$ observed in every case are a simple manifestation of the curvature apparent in the rotational diagrams.  

We have obtained two-component fits to the rotational diagrams, in which a least squares fit was obtained for a combination of warm and hot gas components, assumed to be in LTE except with non-LTE ortho-to-para ratios, 
with temperatures  $T_w$ and $T_h$ (typically found to be $\sim 400$~K and $\sim 1000$~K, respectively).   
The solid lines in Figure 11
show the two-component fits to the 
rotational diagrams, while the points indicate the measured values.  The two-component fits 
provide an excellent fit to the data, although the solution is clearly non-unique.  Indeed, there is no specific 
evidence for a bimodal temperature distribution, and the data could be fit equally 
well by models that invoke a continuous range of gas temperatures.
The assumption of LTE is clearly valid for the lower rotational states, but the data would allow 
modest departures from LTE for the highest states that we have observed.  
Indeed, calculations by Le Bourlot et al.\ (1999) indicate that the critical density for $J=9$ (at which the probabilities of collisional and radiative de-excitation are equal) is $\sim 4 \times 10^5 \rm \, cm^{-3}$, a factor of several larger than the densities inferred from observations of CO in some of these sources (see discussion below).
Departures from LTE, if present, would
be an increasing function of the rotational quantum number, $J$, and 
would therefore require higher gas temperatures than the values inferred under the assumption of LTE.

The best-fit temperatures, computed under the assumption of LTE, 
together with the beam-averaged column densities ($N_w$ and $N_h$) and the H$_2$ ortho-to-para ratios (OPR$_w$ and OPR$_h$), are listed in Table 3.  Table 3 also shows the sum of the $N(4)$ though $N(9)$ column densities, $N_{4-9}$, all of which are derived directly from observations with the Short-Low module; not surprising, $N_{4-9}$ falls a factor of several below $(N_w + N_h)$ because it only includes states with $J \ge 4$. 
The estimated gas temperatures, $T_w$ and $T_h$, are characteristic of low-velocity 
nondissociative shocks (which are of ``C"-type in the designation of Draine (1980); see also Appendix B).  By means of a simple analytic treatment of energy conservation in such shocks (Appendix B), constrained by the results of the detailed shock models of Kaufman \& Neufeld (1996; hereafter KN96), we find the characteristic gas temperature within such shocks to be given by 
$$T_s = 375\, b^{-0.36} v_{s6}^{1.35}\, \rm K$$
(eqn.\ B8), where $10\,v_{s6}\, \rm km \,s^{-1}$ is the shock velocity and 
$b (n_0 / {\rm cm}^{-3})^{1/2} \, \rm \mu G$ is the assumed preshock magnetic field for a shock propagating in material of preshock H$_2$ density $n_0$.  Given our standard assumption about the magnetic field, $b=1$, the warm and hot gas components typically correspond to shock velocities of $\sim 10$ and $\sim 20$ $\rm km \, s^{-1}$ respectively.  The H$_2$ rotational diagrams predicted by KN96 and other theoretical shock models (e.g. Wilgenbus et al.\ 2000) all exhibit less curvature than our observed rotational diagrams, implying the presence of a range of shock velocities within each of the circular apertures demarked in Figure 7.

The preshock H$_2$ density, $n_0$, is less well constrained by the H$_2$ line fluxes.  The best estimates for the density of the warm shocked gas have been obtained by observations of high-lying rotational lines of CO, for which non-LTE line ratios provide a valuable probe.  Based upon {\it ISO} observations of such transitions toward HH54, Liseau et al.\ (1996) estimated the H$_2$ density in the warm shocked gas
as $1 - 4 \times 10^5 \rm \, cm^{-3}$.  Since most of the high-$J$ CO emission occurs before the compression of the neutral species becomes large, the corresponding preshock density is only a factor $\sim 1.5$ smaller\footnote{Liseau et al.\ significantly overestimated the compression of the neutral material, thereby obtaining a significant underestimate of the preshock density required to match the CO line ratios.} (Appendix B, eqn.\ B6): $n_0 = 6 - 25 \times 10^4 \rm \, cm^{-3}$.  Similar  values were obtained for HH7--11, based upon measurements of CO line ratios (Molinari et al.\ 2000), although measurements of the CO/H$_2$ line ratios argued for smaller values ($n_0 \sim 10^4 \rm \, cm^{-3}$) toward HH7.

The observed column density of shocked H$_2$ is predicted to be
$$N_s = 4 \times 10^{20} \, b\, n_{04}^{0.5}  \, v_{s6}^{-0.75}\, \Phi_s,$$
(Appendix B, eqn.\ B7),
where $\Phi_s$ is a geometric factor defined to be the ratio of the projected beam size to the
surface area of the shock.
In principle, the quantity $\Phi_s$ can be either larger or smaller than unity.  For a planar shock with a beam covering factor of unity that is viewed at angle $\theta$ to the shock normal, the observed value of $\Phi_s$ is ${\rm sec}\,\theta \ge 1$.  On the other hand, $\Phi_s$ can be smaller than unity when the beam is incompletely covered by shocked material.  For our two component fits to the circular apertures presented in Table 3, the values of $\Phi_s$ required for the warm and hot components are typically $0.2\,n_{04}^{-0.5}$ and $0.05\,n_{04}^{-0.5}$, respectively, implying that the shocked material has a covering factor considerably smaller than unity and possesses substructure that is not resolved by these observations.  In HH7, the subarcsecond resolution maps of near-IR H$_2$ vibrational emissions presented by Noreiga-Crespo et al.\ (2002) clearly indicate that $\Phi_s \ll 1$ for observations with the IRS instrument.

The spectral line maps obtained for HH54 and HH7 also permit us to obtain a rotational diagram at each spatial position, and thereby to derive detailed maps of the gas temperature, ortho-to-para ratio, and column density of the shocked gas.  
While the average spectra presented in Figure 8 are of sufficient signal-to-noise ratio (SNR) to permit the two-component fits described above, the lower SNR obtained in a single spectrum does not warrant a two-component fit to the rotational diagram.  Accordingly, in constructing maps of the gas temperature, we obtained a least-squares fit to the $N(4)$ through $N(7)$ column densities (all measured with the SL module) for a {\it single} gas component with a temperature and H$_2$ ortho-to-para ratio, maps of which are presented in Figure 12 (middle and lower panels respectively).  Here, the white contours show the H$_2$ S(3) line intensity, and regions of low SNR have been masked and appear in black.  The upper panel in Figure 12 shows the sum of the $N(4)$ though $N(9)$ column densities, $N_{4-9}$. Figure 12 shows that the fitted gas temperature is remarkably constant over the mapped region, lying generally within the range $900 \pm 200$~K, and indicating that a similar admixture of shock velocities is present along each line-of-sight.  The ortho-to-para ratio, however, shows substantial variations -- as noted previously by Wilgenbus et al.\ (2001) -- and is discussed further in \S 4.4

\subsection{Upper limits on the H$_2$O abundance}

The increase in temperature and density behind a shock front
can alter significantly the composition of the gas in its wake.  
Specifically, chemical reactions not favored at temperatures
characteristic of the preshock quiescent material (i.e., 
$T\,\simlt\,$50~K) can proceed rapidly in the warm post-shocked
gas.  One such example are the reactions that lead to the 
production of gas-phase \water.  In the cold preshock gas, 
cosmic-ray driven ion-neutral reactions are capable of converting 
O and H into water yielding a peak water abundance relative to H$_2$, 
$x({\rm H_2O})\,\equiv \,n({\rm H_2O})/n({\rm H_2})$, of 
$\sim\,$10$^{-6}$ in approximately 10$^4$ years (Bergin et al.\ 1998).
When the gas temperature exceeds $\sim\,$300~K, a series of neutral-neutral
reactions, $\rm H_2 + O \rightarrow OH + H$ and $\rm H_2 + OH 
\rightarrow H_2O + H$, dominate and chemical models (e.g., Elitzur 
\& de Jong 1978; Neufeld et al.\ 1995) predict that most of the 
gas-phase oxygen not bound as CO will be processed into water vapor.
These reactions are rapid in warm gas, producing 
$x({\rm H_2O})\,\sim\,$2$\times$10$^{-4}$ in less
than a few hundred years (Bergin et al.\ 1998),
but are negligibly slow at low temperatures because they possess
significant activation barriers.

Observations of both HH54 and HH7 have been conducted using ESA's
{\em Infrared Space Observatory (ISO)}.  {\it ISO}'s Long Wavelength 
Spectrometer (LWS) was used in its grating mode to acquire low
resolution ($\lambda/\Delta\lambda \sim\,$200) spectra between 43 and
197\um\ within an 80$^{\prime\prime}$ diameter beam.  Toward HH54,
Liseau \etal\ (1996) detected a number of low-lying (i.e., 
$E/k\,\simlt\,$300~K above the ground state) \water\ and OH lines, along
with CO emission lines from the rotational upper level
$J\,=\,$14 to 19.  These data are best fit by a model that assumes
collisional excitation behind a 10-15~km~s$^{-1}$ nondissociative
$C$-type shock and $x({\rm H_2O})\,=\,$10$^{-5}$.

Toward HH7, Molinari et al.\ (2000) find evidence for the presence of both 
fast dissociative $J$-type and slower $C$-type shocks.  In particular,
Molinari et al.\ find
that a 15-20 km~s$^{-1}$ $C$-type shock accounts well for the
observed strengths of the CO $J\,=\,$14--13 to $J\,=\,$17--16 transitions
plus the H$_2$ (0-0) S(0)--S(5) transitions, which were measured with the
{\it ISO} Short Wavelength Spectrometer.  Assuming that the {\it ISO}-detected 
\water\ 179.5\um\ 2$_{12}-$1$_{01}$ and 174.6\um\ 3$_{03}-$2$_{12}$ lines
arise predominately from behind the same $C$-type shock, Molinari
\etal\ derive $x({\rm H_2O})\,\leq\,$6$\times$10$^{-6}$.  Molinari et al.\
suggest that the lower-than-expected post-shock \water\ abundance is due
to the rapid condensation of water onto dust grains in the post-shocked
gas.  A similar effect is observed toward IC$\:$443, in which a low
water abundance 
(2$\times$10$^{-8}\,<\,x({\rm H_2O})\,<\,$3$\times$10$^{-7}$) is
inferred\footnote{Snell et al.\ expressed their findings in terms
of the ratio $n({\rm o-H_2O})/n({\rm CO})$.  Here we make the assumption
that $n({\rm CO})/n({\rm H_2}) = 10^{-4}$.}
in the presence of both $C$- and $J$-type shocks (Snell et al.\ 2005).
Snell et al.\ ascribe the low water abundance to a combination of two effects.  
First, more than half of the preshock oxygen not in CO remains 
trapped as water-ice on grain surfaces despite the passage of a weak 
$C$-type shock, suppressing the production of gas-phase \water\ in the
post-shocked gas.  Second, ultraviolet radiation generated by the nearby
fast $J$-type shock photodissociates much of the gas-phase water 
produced behind the $C$-type shock.

Eight H$_2$O transitions with upper state energies $E_U/k < 1000$~K lie
within the wavelength range accessible to {\it Spitzer}/IRS.
Prior to the measurements reported here, 
none of these transitions had been observationally investigated 
toward either HH54 or HH7.
To assess the extent to which the upper limits to the \water\ line strengths
set here constrain the water abundance, the equilibrium level
populations of all ortho and para rotational levels of the ground
H$_2^{\;16}$O vibrational state with energies {\em E/k} up to 7700~K
have been calculated using an escape probability method described by
Neufeld \& Melnick (1991).  Only two updates have been made in the calculations
undertaken here: (1) for collisionally induced transitions among the lowest
45 levels of ortho- and para-water, the rate coefficients calculated by
Green, Maluendes, \& McLean (1993) for He-H$_2$O collisions were used
(multiplied by 1.348 to account for the different reduced mass when H$_2$
is the collision partner), and (2) the expression for the photon escape
probability used by Neufeld \& Kaufman (1993) was adopted.

We obtain level populations for a plane-parallel emitting region in which
the limit of large velocity gradient applies.   A two-temperature component
model is used, consistent with the results of our H$_2$ measurements (which
are summarized in Table~3).  In addition, preshock densities of 10$^4$
and 10$^5$~\cmc\ are considered.  The H$_2$ column densities and
velocity gradients are assumed to be those consistent with the simple
shock model presented in Appendix~B, i.e.:

\vspace{2.5mm}

\begin{displaymath}
N(H_2) = 7.9 \times 10^{20}\;\left[\frac{n(H_2)}{10^5}\right]^{0.5}\,
\left[\frac{T_{gas}}{1000}\right]^{-0.555}~~~{\rm cm^{-2}}
\end{displaymath}

\vspace{-2mm}

and

\vspace{-2mm}

\begin{displaymath}
\frac{dv}{dz} = 1.6 \times 10^4\;\left[\frac{n(H_2)}{10^5}\right]^{0.5}\,
\left[\frac{T_{gas}}{1000}\right]^{1.30}~~~{\rm km~s^{-1}~pc^{-1}},
\end{displaymath}

\vspace{2.5mm}
\noindent where $n(H_2)$ is the preshock gas density and $T_{gas}$ is the gas
temperature.  Finally, we assume that the fraction of the beam filled by
the warm and hot \water\ is given by the ratio of the measured values of
$N({\rm H}_2)$ (given in Table~3) to those derived above.  For these
conditions, the strongest water lines within
the {\em Spitzer}/IRS wavelength range are predicted to be the 29.837\um\
7$_{25}$-6$_{16}$ and 30.899\um\ 6$_{34}$-5$_{05}$ ortho-\water\
transitions.  Table~4 lists the {\em Spitzer}-measured 3$\sigma$ upper
limits to the line intensities for each transition along with the 
derived upper limits to the gas-phase water abundance.

Several things are apparent.  First, the upper limits to the \water\
abundance given in Table~4 are higher than, but nonetheless
consistent with the water abundances derived
previously from the {\it ISO} data.   In particular, for a preshock density of 
10$^5$~\cmc\ -- favored by the H$_2$ data toward HH54 (Liseau \etal\
1996) -- the inferred abundance limits
are less than would be expected from the full conversion of elemental
oxygen not locked in CO into \water.  Unfortunately, the present 
29.837 and 30.899\um\ line flux limits are insufficient to allow us
to comment further upon the previous HH54 models.
Second, the \water\ 174 and 179\um\ line fluxes measured by Liseau \etal\
provide an additional constraint on $x({\rm H_2O})$ within the warm and hot
components discussed here.  In particular, the {\it ISO}-measured flux in the
\water\ 174\um\ line, $\sim\,$3$\times$10$^{-20}$ W~cm$^{-2}$, requires that
$x({\rm H_2O})\,<\,$2$\times$10$^{-5}$ in the warm and hot H$_2$ components or
these components will produce a 174\um\ line flux in excess of that measured.
This result is in agreement with the water abundance determined by
Liseau et al.\ of 10$^{-5}$ and suggests that the size
of the \water\ 174 and 179\um\ emitting region is larger than that inferred for the
{\em Spitzer}-measured H$_2$ emission (or that assumed for the mid-infrared \water\
emission).  This is plausible since the 174 and 179\um\ lines probe cooler gas 
than either the H$_2$ or mid-infrared \water\ lines.

Toward HH7, Molinari et al.\
estimate the preshock density to be $\sim\,$10$^4$ cm$^{-3}$.  
For this density, rather high H$_2$O abundances are
required to reproduce the 29.837 and 30.899\um\ line intensity limits.
However, the strength of the \water\ 179\um\ line observed by Molinari et al.\
does set a more stringent upper limit of $x({\rm H_2O})\,<\,$6$\times$10$^{-4}$ within
the warm and hot components measured here.  Unfortunately, this limit remains
significantly greater than the water abundance inferred from the {\it ISO}
measurements and, thus, does not permit a test of their model.

\subsection{The H$_2$ ortho-to-para ratio}

The H$_2$ ortho-to-para ratios (Table 3) derived for the two-component fits, OPR$_w$ and OPR$_h$, are uniformly less than the local thermodynamic ratio, $\rm OPR_{\rm LTE}\sim 3$ at the temperatures of relevance.  This non-equilibrium behavior, first observed in the atmospheres of giant planets (e.g. Conrath \& Gierasch 1983), was noted in {\it ISO} observations of HH54 (NMH98; Wilgenbus et al.\ 2001) and subsequently observed in several other interstellar regions (e.g.\ Fuente et al.\ 1999, Rodr{\'{\i}}guez-Fern{\'a}ndez et al.\ 2000).  As discussed in NMH98, the H$_2$ ortho-to-para ratio equilibrates very slowly, exhibiting ``memory effects" in which the OPR can reflect the temperature of a previous epoch in its thermal history\footnote{Such effects are a well-known problem in the industrial production and storage of liquid hydrogen (LH2).  The rapid refrigeration of molecular hydrogen leads to the production of LH2 with an ortho-to-para ratio $\sim 3$, reflecting the ``frozen-in" equilibrium value at room temperature, and greatly exceeding the value of 0.02 obtained at equilibrium at 21~K, the boiling point of LH2.  Equilibration occurs over a timescale of 6.5 days, resulting in the eventual release of a total heat content that exceeds the latent heat of vaporization of LH2: without continued refrigeration, the LH2 simply boils away!  The solution to this problem involves the introduction of a paramagnetic catalyst to facilitate ortho-to-para conversion during the original liquification process and thereby prevent the room temperature OPR from being frozen-in.}.  Thus NMH98 argued that the OPR of $1.0 \pm 0.4$ observed toward HH54 indicated (1) that the warm ($T \sim 650$~K) shocked H$_2$ that they observed in HH54 must have previously acquired an OPR characteristic of equilibrium at a temperature $\le 90$~K and (2) that the observed gas has not been warm long enough for a high-temperature OPR to have been established.

The equilibration of ortho- and para-H$_2$ has been considered in the detailed shock models of Timmermann (1998) and Wilgenbus et al.\ (2000).  Both studies concluded that reactive collisions with atomic hydrogen are the dominant para-to-ortho conversion process in non-dissociative molecular shocks:
$$\rm \hbox{para-H}_2 + H \rightarrow H + 
\hbox{ortho-H}_2 \eqno(R1)$$
Because reaction (R1) has a substantial activitation energy barrier ($\Delta E_A/k \sim 3900$~K), the efficiency of para-to-ortho conversion is a strong function of temperature (and, consequently, shock velocity).  This strong temperature dependence is entirely consistent with the results in Table 3, which show that for each region the hot gas component has a much higher OPR than the warm gas component.  This behavior is also apparent in the H$_2$ rotational diagrams (Figure 11), in which the departures from LTE (i.e.\ the degree of zigzag) are plainly larger for small $J$ than for large $J$.

In Figure 13, the ortho-to para ratios, OPR$_w$ and OPR$_h$, for the two-component fits in Table 3, are plotted as a function of the fitted gas temperatures, $T_w$ and $T_h$ (open triangles).  We have also obtained two-component fits to the gas temperature and OPR elsewhere in the mapped regions.  Averaging the observed spectra in a set of $5^{\prime\prime} \times 5^{\prime\prime}$ subregions to attain adequate SNR to yield meaningful two-component fits, we determined 
OPR$_w$, OPR$_h$, $T_w$ and $T_h$ estimates for each such subregion.  Green and red squares in Figure 13 indicate respectively the ($T_w, $OPR$_w$) and ($T_h$, OPR$_h$) values obtained in subregions within the HH54 and HH7 maps.  Green and red crosses apply to the discontiguous regions northwest of HH7 (Figure 12; these are close to HH8).  The orange curve denotes the ortho-to-para ratio expected in LTE.

The green and red symbols in Figure 13 confirm that the ortho-to-para ratio tends to increase with gas temperature.  The black curves show the expected behavior for a simple model in which all the observed gas started with the same initial ortho-to-para ratio -- OPR$_0 = 0.4$ (corresponding to equilibrium at temperature 55~K) for HH54 and OPR$_0 = 0.25$ (corresponding to equilibrium at temperature 48~K) for HH7 -- and has been at the currently-observed temperature $T$ for the same time period, $\tau$.  In that case, the current OPR is given by 
$${\rm OPR(\tau) \over 1+OPR(\tau)} =  {\rm OPR_0 \over 1+OPR_0} \, e^{-n({\rm H})\, \tau \, k_{\rm PO}} +  {\rm OPR_{LTE} \over 1+OPR_{LTE}} \, \biggl( 1- e^{-n(H)\, \tau \, k_{\rm PO} }\, \biggr),$$
where $n({\rm H})$ is the atomic hydrogen density and $k_{\rm PO}$ is the rate coefficient for para-to-ortho conversion, estimated as $8 \times 10^{-11} \exp(-3900/T) \rm \, cm^3 \, s^{-1}$ by Schofield et al.\ (1967), based upon the laboratory experiments of Schulz \& Le Roy (1965).
The black curve applies to the case $n_2 ({\rm H}) \tau = 120$~yr, where $n_2 ({\rm H}) = n({\rm H})/10^2 \rm  \, cm^{-3}$. If, following NMH98, we assume that H atoms are produced primarily by reaction of O with H$_2$ to form OH, followed by reactions of OH with H$_2$ to form water, we expect an $n({\rm H})/n({\rm H}_2)$ ratio $\sim 10^{-3}$, corresponding to $n_2 ({\rm H}) \sim 1$ for a preshock H$_2$ density $n_0 = 10^5 \rm \, cm^{-3}$.
The steep rise in the black curve at $T \sim 1000$~K is a consequence of the strong temperature-dependence of $k_{\rm PO}$.  To within the likely errors, all the points lie between the black and orange curves in Figure 13.  This behavior then implies that all the shocked gas had an initial OPR between 0.4 and 3 and has been warm for at least $150\,n_2({\rm H})^{-1}$~yr.  The foregoing analysis is premised on the assumption that the rotational temperature inferred for the hot gas component is equal to the kinetic temperature.  As discussed in \S4.2, above, subthermal excitation of the S(7) transition is allowed by the data and would mean that the actual gas temperature is higher than the rotational temperature for the hot component, $T_h$.  Such effects, in turn, would decrease the required timescale for ortho-para conversion.
 
For a postshock atomic H abundance $\sim 10^{-3}$, the timescale of $150\,n_2({\rm H})^{-1}$~yr inferred above is broadly consistent with the time period for which gas temperature is elevated upon passing through a nondissociative shock.  For $n_0 = 10^5 \rm \, cm^{-3}$ and $v_s = 15\,\rm  km\,s^{-1}$, the simple treatment of Appendix B implies a shock column density, $N_s \sim 10^{21} b\, \lambda_{in}\rm cm^{-3}$, corresponding to a timescale $N_s/(n_0 \, v_s) =  200\,b\, \lambda_{in}$~yr.  However, the OPR values plotted in Figure 13 are entirely inconsistent with a shock model with a large H abundance, as has been invoked by Smith et al.\ (2003), who proposed an $n({\rm H})/n({\rm H}_2)$ ratio of 3 for the bow shock in HH7.  Such a large H abundance would lead to very rapid para-to-ortho conversion, producing OPR values significantly in excess of those observed here.

We may also compare the results plotted in Figure 13 with the predictions of detailed theoretical models for para-to-ortho conversion in non-dissociative molecular shocks.  The cyan loci in Figure 13 represent the predictions of Wilgenbus et al.\ (2000) for a shock propagating in gas with an initial OPR of 0.01. The loci correspond to preshock H$_2$ densities of $10^3$, $10^4$, 10$^5$ and $10^6\, \rm cm^{-3}$, respectively from left to right, and the asterisks apply to shock velocities of 30 and 40 $\rm km \, s^{-1}$ ($10^3 \, \rm cm^{-3}$ case, from lower left to upper right); 20, 30, and 40 $\rm km \, s^{-1}$ ($10^4$ and $10^5 \, \rm cm^{-3}$ cases, from lower left to upper right); or 10, 15, 20, 25, and 30 $\rm km \, s^{-1}$ ($10^6 \, \rm cm^{-3}$ case, again from lower left to upper right).  As in the case of the toy model (black curve), the curves for 10$^5$ or $10^6 \, \rm cm^{-3}$ are a reasonable approximation to the lower envelope of the observed data points.  The absence of sight-lines with high temperature and low OPR is consistent with shock models in which para-to-ortho conversion is driven by reactive collisions with atomic hydrogen.  On the other hand, the presence of sight-lines with {\it low} temperature and {\it high} OPR suggests that some of the shocked material has a high {\it initial} OPR.  Such material is clearly present for the sight-lines represented by the crosses in Figure 13.  These regions lie behind the ``working surface" of the jet in HH7--11; they may, perhaps, have been heated by relatively fast shocks during a previous epoch of shock activity, thereby acquiring a larger OPR, and are currently being reheated by a slower shock.  The bow shock in HH7, by contrast, appears to be propagating into fresh material in which the OPR is $\simlt 0.25$, corresponding to equilibrium at temperature $\sim 50$~K.

\section{Summary}

We have carried out spectroscopic observations toward the Herbig-Haro objects HH7--11 and HH54 over the $5.2 - 37\rm\,\mu m$ region using the Infrared Spectrograph of the {\it Spitzer Space Telescope}, and have thereby mapped the S(0) -- S(7) pure rotational lines of molecular hydrogen, together with emissions in fine structure transitions of Ne$^+$, Si$^+$, S, and Fe$^+$.  

The primary results of our study are:

1.\ The H$_2$ rotational emissions indicate the presence of warm gas with a mixture of temperatures in the range $400 - 1200$~K, consistent with the expected temperature behind nondissociative shocks of velocity $\rm \sim 10 - 20\, km\, s^{-1}$.

2.\ The fine structure emissions are spatially offset from the H$_2$ emissions, lying closer to the source of the outflow that drives the shocks in these sources.  They originate in faster shocks -- of velocity $\sim 35 - 90\, \rm km \, s^{-1}$ -- that are dissociative and ionizing.  Such shocks likely result from the deceleration of the outflow, and are offset from a slower non-dissociative bow shock that is driven into the ambient molecular medium.

3.\ Maps of the H$_2$ line ratios reveal little spatial variation in the typical admixture of gas temperatures in the mapped regions, but show that the H$_2$ ortho-to-para ratio is quite variable, typically falling substantially below the equilibrium value of 3 attained at the measured gas temperatures.  

4.\ The non-equilibrium ortho-to-para ratios are characteristic of temperatures as low as $\sim 50$~K, and are a remnant of an earlier epoch, before the gas temperature was elevated by the passage of a shock.  Correlations between the gas temperature and H$_2$ ortho-to-para ratio show that ortho-to-para ratios $< 0.8$ are attained only at gas temperatures below $\sim 900$~K; this behavior is consistent with theoretical models in which the conversion of para- to ortho-H$_2$ behind the shock is driven by reactive collisions with atomic hydrogen, a process which possesses a substantial activation energy barrier ($E_A/k \sim 4000$~K) and is therefore very inefficient at low temperature.  

5.\ The lowest observed ortho-to-para ratios of only $\sim 0.25$ suggest that the shocks in HH54 and HH7 are propagating into cold clouds of temperature $\simlt 50$~K in which the H$_2$ ortho-to-para ratio is close to equilibrium.

6.\ Upper limits on the 29.837\um\
7$_{25}$-6$_{16}$ and 30.899\um\ 6$_{34}$-5$_{05}$ transitions of ortho-\water\ place upper limits on the \water\
abundance that are higher than -- and therefore consistent with -- abundances derived
previously from {\it ISO} observations of transitions at longer wavelength.  

\acknowledgments

This work is based on observations made with the {\it Spitzer Space Telescope}, which is operated by the Jet Propulsion Laboratory, California Institute of Technology, under a NASA contract.
D.A.N.\ and P.S.\ gratefully acknowledge the support of grant NAG5-13114 from NASA's Long Term Space Astrophysics (LTSA) Research Program. 

\vfill\eject
\centerline{\bf Appendix A: data reduction procedures}

Our data reduction procedure involved the following steps:

1) {\it Removal of bad pixels in the ``Basic Calibrated Data (BCD)" files that are generated by the Spitzer Science Center for the Short-High and Long-High modules.}  Bad pixels were defined as those flagged in either the SSC ``bmask" file or in a ``grand rogue mask" generated at University of Rochester from an analysis of dark fields acquired at the start of each IRS campaign.  As a rogue might not
appear in every campaign dark field, we assumed that any pixel identified
as a rogue in ANY campaign would be corrected for all campaigns; thus, the
grand rogue mask contains ~ 25$\%$ more rogue pixels than a single campaign
rogue mask.  They were removed with the ``imclean" routine, which interpolates between nearby good pixels.

2) {\it Extraction of the individual spectra at each position along the slit.}  These spectral extractions were accomplished using an augmentation to the SMART software package that has been developed at University of Rochester.  This augmentation generates a spectrum for each resolution element along the
slit, resulting in a series of spectra at defined positions in R.A. and
declination.  The spectra were trimmed to eliminate overlap between different orders of the IRS.

3) {\it Regridding of the spectra onto a grid that is regular in R.A. and declination.}  Because there are positions that are multiply observed with more than one slit position, this regridding sometimes involves the averaging of spectra obtained at nearly identical positions.

4) {\it Application of the slit loss correction function (SLCF).}  Because the SSC-provided data are flux-calibrated using point source calibrators, a downwards correction must be provided to obtain intensities for extended sources, the correction factor (i.e.\ the ``SLCF") being the fraction of the point source flux that falls within the slit width.  We used the SLCFs kindly provided by J.~D.~Smith from an analysis of the instrumental point spread function.

5) {\it Determination of line and continuum fluxes at each spatial position within the mapped region.}  Here, we used a Levenberg-Marquardt (L-M) fitting routine simultaneously to fit a Gaussian line and second order polynomial baseline at each position.  Line fits were obtained for a series of spectral lines identified in the co-averaged spectra for each field.  In fitting the co-averaged spectra, which are of high signal-to-noise ratio (SNR), the line centroids and widths were treated as free parameters to be optimized in the L-M fit.  The resultant best-fit centroids and widths were then used as {\it fixed} parameters in the L-M fits to the lower SNR spectra obtained at individual spatial positions; for these, only the line intensities and continuum parameters were optimized.  This procedure is justified by the fact that the expected Doppler shifts associated with gas motions in the source are very small compared to the spectral resolution, $\lambda/\Delta \lambda$ being only $\sim 60$ and $\sim 600$ respectively for the low (SL) and high (SH and LH) resolution modules.

The data reduction procedure was accomplished using the SMART software package and a set of associated IDL procedures that we have written.  The contour maps presented in Figures 1 -- 7 were also created using IDL procedures.

To verify our data reduction procedure, we have conducted several tests.  These include:

1) {\it Continuum mapping of SVS13.}  Figure 14 shows contour maps obtained for several continuum wavelengths for HH7--11 Field 1 (see Table 1), which contains the strong continuum source SVS13 at its center (i.e. at position (0,0) in Figure 14).  The intensity maps shown in Figure 14 use logarithmic contours spaced by a factor $2^{1/2}$, with the second-highest contour corresponding to an intensity one-half the maximum.  To the extent that SVS13 can be regarded as an unresolved point source at these wavelengths, the maps in Figure 14 represent the beam profile resulting at the end of the entire data reduction procedure.  These maps allow the astrometric accuracy and the half-power-beam width (HPBW) to be estimated for the final data product.  Table 5 shows the results of Gaussian fits to the spatial profiles, and indicates that the astrometric accuracy is better than $\sim 2 ^{\prime \prime}$ in the final data product.  The map at 5.8 $\mu$m clearly exhibits a sixfold rotational symmetry at intensities levels less than $\sim 3\%$ of the maximum, a behavior that is also evident in Infrared Array Camera (IRAC) observations (Hora et al.\ 2004) and reflects the diffraction pattern produced by the telescope.   

2) {\it Spectral extraction of SVS13 and comparison with {\it Infrared Space Observatory} (ISO) observations.}  As a check of the flux calibration, we have extracted spectra of the strong continuum source SVS13, obtaining a co-average from the entire map of HH7--11 Field 1.  For comparison, we have obtained an archival spectrum of SVS13 from the {\it ISO} Short Wavelength Spectrometer (SWS), co-averaging multiple scans and removing bad pixels with the use of the ISAP software package.  Figure 15 compares the spectrum obtained with {\it Spitzer} (heavy black line) with that obtained with {\it ISO} (red, blue, green and magenta lines, different colors being used to show the different SWS orders).  Overall, the agreement is very good, with the absolute fluxes typically agreeing to within $\sim 15\%$.   In the spectral region longward of 27.5 $\mu$m, the fluxes obtained with {\it ISO}/SWS exceed those measured by {\it Spitzer} by as much as $25 \%$. However, a significant order mismatch at 27.5 $\mu$m (between detectors 11 and 13) in the {\it ISO} spectrum suggests that the {\it ISO}-determined flux is an overestimate longward of 27.5 $\mu$m.

\vskip 0.3 true in
\centerline{\bf Appendix B: simple analytic treatment of nondissociative C-type shocks}

Here we present a simple analytic treatment of nondissociative C-type shocks, which is limited in its applicability to steady-state shocks with plane parallel geometry.
Detailed models of planar nondissociative C-type shocks (e.g.\ Kaufman \& Neufeld 1996, hereafter KN96; Wilgenbus et al.\ 2000), predict line strengths for the H$_2$ pure rotational transitions that yield rotational diagrams with remarkably little curvature.  This behavior suggests that the warm gas behind such shocks can be approximated as an isothermal, isobaric slab.  In this appendix, we show that such a description yields predictions for the emission line spectrum of the shocked gas that are in good agreement with those of detailed models, and we derive analytic expressions for the appropriate slab thickness, density, temperature and velocity gradient as a function of the preshock H$_2$ density, $n_0 =  n_{05} \,10^5 \, \rm cm\,^{-3}$, shock velocity, $v_s = 10\, v_{s6}\rm \, km \, s^{-1}$, and preshock magnetic field,  $B_0 = b\,(n_0 /\rm cm\,^{-3})^{1/2} \,\mu G$.

Nondissociative C-type shocks have been discussed extensively in the literature (e.g. Draine \& McKee 1993, and references therein) and will not be reviewed here.  They were originally defined by Draine (1980), who distinguished between J- (or ``jump")-type shocks in which the fluid velocity changes suddenly and the neutral and ionized species share a common velocity, and C- or (``continuous")-type shocks in which the fluid velocity varies slowly.  The basic physics of C-type shocks had been discussed previously by Mullan (1971), who recognized ion-neutral drift to be an essential feature of magnetic shocks propagating in weakly-ionized media.  Ions entering the shock, being strongly coupled to the magnetic field, are decelerated fairly rapidly and quickly reach their maximum compression.  Neutral species entering the shock, by contrast, are uncoupled to the magnetic field and are slowly decelerated in collisions with the ions through which they drift.  Thus the maximum heating rate within the shocked material occurs where the compression of the ions is large and that of the neutral species is small.  Accordingly, we make the following simplifying approximations:

1) The neutral fluid is at a constant compression, $C_n = n({\rm H}_2)/n_{0}$, which is slightly greater than unity; the exact value is an adjustable parameter to be determined below by a comparison with the line flux predictions of the KN96 shock models.

2) The ionized fluid is at a constant compression equal to the maximum that is
attained behind the shock
$$C_i = n_i / n_{i0} = 2\,M_A^{1/2} = 7.7 v_{s6} / b  ,\eqno(B1)$$
where $M_A = v_{s6}/(0.184 b)$ is the Alfven Mach number and $n_{i0}$ is the preshock ion density.
Additional ionization within the shock (``self-ionization") is neglected.

3) The thickness of emitting slab is given by
$$t = k_1 L_{in} (n_0, v_s) / C_i, \eqno(B2)$$
where $L_{in} (n[{\rm H}_2], v_d)$ is the ion-neutral coupling length (KN96) for H$_2$ density $n({\rm H}_2)$ and ion-neutral drift velocity $v_d$, and $k_1$ is a dimensionless constant of order unity, the exact value of which is an adjustable parameter to be determined below by a comparison with the line flux predictions of the KN96 shock models.

4) The slab is at constant temperature, $T$, determined such that the radiative output is equal to the mechanical luminosity, implying
$$n_0^2C_n^2 t \Lambda (T, n_0 C_n) = \onehalf \mu n_0 v_s^3 (1 - [2M_A]^{-2}) \sim \onehalf \mu n_0 v_s^3, \eqno(B3)$$
where $\Lambda(T, n[{\rm H}_2])$ is the cooling rate coefficient for temperature, $T=10^3\,T_3\,$K, and H$_2$ density, $n({\rm H}_2) = 10^5 \, n_5 \,\rm cm^{-3}$.

5) The velocity gradient is $dv_z/dz = k_2 v_s/t,$
where $k_2$ is a dimensionless constant of order unity, the exact value of which is an adjustable parameter to be determined below by a comparison with the line flux predictions of the KN96 shock models.

6) We adopt as our standard assumptions an ion-neutral coupling length 
$$L_{in}^0 = 3.5 \times 10^{16} n_5^{-0.5} v_{d6}^{0.25}\,\rm cm, \eqno(B4)$$
based upon a fit to results of KN96 (obtained for the case where polycyclic aromatic hydrocarbons (PAHs) are assumed to be present), and a cooling rate coefficient
$$\Lambda^0(T,n[{\rm H}_2]) = 1.4 \times 10^{-26} n_5^{-0.5} T_3^{2.8}\,\rm erg \, cm^3 \,s^{-1}  \eqno(B5)$$
based upon a fit to the results of Neufeld et al.\ (1995).  The superscripts on $L_{in}^0$ and 
$\Lambda^0$ are used to denote the ``standard values" adopted for our treatment.

Assumptions (1) -- (6) above define the required properties of the isothermal, isobaric slab, once the three parameters $C_n$, $k_1$, and $k_2$ have been specified.  In determining the best-fit values for these parameters, we have compared the emission predicted for the slab model with the detailed predictions of KN96.   Here we assumed fixed CO and H$_2$O abundances, corresponding to the assumption that CO and H$_2$O account for all the O and C nuclei in the gas-phase.  Because the O and C abundances assumed by KN96 were somewhat larger than those adopted by Neufeld et al.\ (1995) in computing the cooling rate coefficient, $\Lambda^0(T,n[{\rm H}_2])$, we adopted $\Lambda(T,n[{\rm H}_2]) = 1.5 \, \Lambda^0(T,n[{\rm H}_2])$ for the purposes of this comparison.  The optimum fit to the KN96-predicted emission line spectrum is obtained for $C_n = 1.5$, $k_1=2$, and $k_2 =2$.

Combining the expressions given above, we find that the slab H$_2$ density is fit best by
$$n({\rm H}_2)=1.5 \times 10^5\,  n_{05}\, \rm cm^{-3}, \eqno(B6)$$
the H$_2$ column density by
$$N({\rm H}_2)=1.37 \times 10^{21}\,  n_{05}^{0.5}v_{s6}^{-0.75}\, b \,(L_{in}/L^0_{in})\, \rm cm^{-2}, \eqno(B7)$$
the gas temperature by
$$T = 375 \, v_{s6}^{1.35} \, b^{-0.36} \, (L_{in}/L^0_{in})^{-0.36}\,(\Lambda/\Lambda^0)^{-0.36} \, \rm K  \eqno(B8)$$
and the velocity gradient by
$$dv/dz = 4500\, n_{05}^{0.5}\, v_{s6}^{1.75} \, b^{-1} \, (L_{in}/L^0_{in})^{-1}\rm \, km\, s^{-1}\,pc^{-1} \eqno(B9)$$

In Figure 16, we compare the H$_2$ S(5)/S(3) line ratio and the total H$_2$, H$_2$O and CO surface luminosities derived for the simple slab model (red) with those obtained in the detailed calculations of KN96 (blue).  In general, the agreement is remarkably good.  Almost invariably, either (a) the shock velocity corresponding to a given flux or flux ratio agrees to within $\sim 5\,\rm km \,s^{-1}$; and/or (b) the predicted line flux or flux ratio for a given shock velocity agrees to within a factor 2.  Often, the agreement is considerably better.  The main exception is for water emissions at low shock velocity ($v_s \le 15\, \rm km \, s^{-1}$), where the detailed KN96 models predict considerably less H$_2$O than the slab models.  This behavior results because -- for the slab models -- we assumed that water vapor accounts for all gas-phase oxygen nuclei not locked in CO, whereas the treatment of oxygen chemistry in KN96 predicts considerably lower water abundances at low shock velocity.

\clearpage

\begin{deluxetable}{llccccc}
\rotate
\tablewidth{0pt}
\tablecaption{Details of the observations}
\tablehead{Source & Date & Module & Observing  & \multicolumn{2}{c}{Map center}
& Map size \\
                             &      &        &  time (s)  & R.A. (J2000) &Dec. (J2000)  & (arcsec)     }
\startdata
HH7--11 Field 1 	&  2004 Feb 15    & SL &  ...          &  3h 29m 3.71s   &  +31d 16$^\prime $ 01.9$^\prime$$^\prime$ &    $58 \times 57$              \\
HH7--11 Field 1 	&  2004 Feb 15    & SH &  ...          &  3h 29m 3.66s   &  +31d 16$^\prime $ 03.6$^\prime$$^\prime$ &    $58 \times 76$               \\
HH7--11 Field 1 	&  2004 Feb 15    & LH & 10243$^1$  &  3h 29m 3.66s   &  +31d 16$^\prime $ 03.6$^\prime$$^\prime$ &         $61 \times 77$         \\
\hline
HH7--11 Field 2 	&  2004 Feb 16    & SL &  ...          &  3h 29m 8.61s   &  +31d 15$^\prime $ 26.6$^\prime$$^\prime$ &      $58 \times 57$            \\
HH7--11 Field 2 	&  2004 Feb 16     & SH &  ...         &  3h 29m 8.56s   &  +31d 15$^\prime $ 28.3$^\prime$$^\prime$ &      $58 \times 76$            \\
HH7--11 Field 2 	&  2004 Feb 16     & LH & 10246$^1$    &  3h 29m 8.56s   &  +31d 15$^\prime $ 28.3$^\prime$$^\prime$ &     $61 \times 77$       \\
\hline
HH54 		        &    2004 Apr 7   & SL & 1133       &  12h 55m 51.52s  &  --76d 56$^\prime $ 19.1$^\prime$$^\prime$ &     $47 \times 64$              \\
HH54   		        &   2004 Apr 7   & SH & 3260        &  12h 55m 50.87s  &  --76d 56$^\prime $ 19.3$^\prime$$^\prime$ &      $47 \times 52$               \\
HH54   		        &   2004 Apr 7   & LH & 931          &  12h 55m 50.82s  &  --76d 56$^\prime $ 18.0$^\prime$$^\prime$ &    $44 \times 67$              \\
\enddata
\tablenotetext{1}{Total for SL, SH, and LH modules}
\end{deluxetable}

\begin{deluxetable}{lrcccc}
\tablewidth{0pt}
\tablecaption{Beam-averaged line intensities}
\tablehead{
Transition & Wavelength &\multicolumn{4}{c}{Intensity ($\rm 10^{-5}\,erg \, cm^{-2} \, s^{-1} \, sr^{-1}$)}\\
\cline{3-6}
& ($\mu$m)&HH54FS & HH54C & HH54E+K & HH7} 
\tabletypesize{\scriptsize}
\startdata
H$_2$ S(0)  $\,\,\,\,J=2-0$ & 28.2188		& 0.52	& 0.44	& 0.47	& 0.78 \\
H$_2$ S(1)  $\,\,\,\,J=3-1$ & 17.0348		& 3.68	& 3.18	& 3.99	& 3.07 \\
H$_2$ S(2)  $\,\,\,\,J=4-2$ & 12.2786 		& 14.1	& 21.4	& 17.2	& 18.2 \\
H$_2$ S(3)  $\,\,\,\,J=5-3$ & 9.6649	 	& 24.7	& 20.7	& 29.5	& 20.6 \\
H$_2$ S(4)  $\,\,\,\,J=6-4$ & 8.0251		& 24.5	& 28.9	& 22.9	& 25.6 \\
H$_2$ S(5)  $\,\,\,\,J=7-5$ & 6.9095		& 33.0	& 24.4	& 32.9	& 29.0 \\
H$_2$ S(6)  $\,\,\,\,J=8-6$ & 6.1086		& 17.0	& 14.0	& 14.3	& 12.0 \\
H$_2$ S(7)  $\,\,\,\,J=9-7$ & 5.5112		& 15.6	& 12.7	& 17.2	& 16.0 \\
\hline
Ne$^+$ $\,\,\,^2P^{\,0}_{1/2} - ^2P^{\,0}_{3/2}$     & 12.8149  & 1.68	& 1.03	& 0.82	& $< 0.36\,\,\,(3\, \sigma)$\\
S $\,\,\,\rm ^3P_{1} -  ^3P_{2}$	             & 25.2490	& 1.75	& 0.75	& 0.75	& 1.73 \\
Si$^+$ $\,\,\,^2P^{\,0}_{3/2} - ^2P^{\,0}_{1/2}$     & 34.8141 	& 6.07	& 2.95	& 3.43	& 4.18 \\
\hline
Fe$^+$ $\,\,\,\rm ^4F_{9/2} - ^6D_{9/2}$             &  5.3403  & $8.6 \pm 0.8^a$  & $6.9 \pm 0.8$  & $2.6 \pm 1.0$	&   \\
Fe$^+$ $\,\,\,\rm ^4F_{7/2} - ^4F_{9/2}$             & 17.9363  & $0.88 \pm 0.18$  & $0.55 \pm 0.08$
& $<0.5\,\,(3\, \sigma)$ & $<0.4\,\,(3\, \sigma)$\\
Fe$^+$ $\,\,\,\rm ^4F_{5/2} - ^4F_{7/2}$             & 24.5186  & $0.23 \pm 0.04$  & $0.16 \pm 0.02$ 
& $0.19 \pm 0.03$ & $<0.12\,\,(3\, \sigma)$ \\
Fe$^+$ $\,\,\,\rm ^6D_{7/2} - ^6D_{9/2}$             & 25.9882  & $3.33$  & $1.69$  & $1.38$ & $1.69$ 
\\
Fe$^+$ $\,\,\,\rm ^6D_{5/2} - ^6D_{7/2}$             & 35.3491  & $0.92 \pm 0.17$  & $0.45 \pm 0.11$  & $0.56 \pm 0.08$   & $0.57 \pm 0.06$  \\ 
\hline

\enddata
\tablenotetext{a}{Error bars, where given, are $1\sigma$ statistical errors.  When error bars are not given, the uncertainty is dominated by systematic errors, which we estimate to be $\le 25\%$ (see Appendix A).}
\end{deluxetable}

\begin{deluxetable}{lcccc}
\tablewidth{0pt}
\tablecaption{H$_2$ parameters}
\tablehead{
\cline{2-5}
&HH54FS & HH54C & HH54E+K & HH7} 
\tabletypesize{\scriptsize}
\startdata
Rotational state &  \multicolumn{4}{c}{log$_{10}\,$(Column density in cm$^{-2}$)}\\
J=2	& 19.51 & 19.44	& 19.47	& 19.70\\
J=3	& 18.94	& 18.88	& 18.98	& 18.89\\
J=4	& 18.62	& 18.81	& 18.71	& 18.77\\
J=5	& 18.25	& 18.17	& 18.33	& 18.26\\
J=6	& 17.70	& 17.77	& 17.67	& 17.75\\
J=7	& 17.40	& 17.27	& 17.40	& 17.36\\
J=8	& 16.78	& 16.69	& 16.70	& 16.65\\
J=9	& 16.45	& 16.36	& 16.49	& 16.48\\
\hline
\multicolumn{4}{l}{Rotational temperatures (K)}\\
$T_{42}$ & 445	& 571	& 501	& 428\\
$T_{53}$ & 729	& 717	& 764	& 779\\
$T_{64}$ & 716	& 650	& 647	& 658\\
$T_{75}$ & 919	& 871	& 852	& 877\\
$T_{86}$ & 986	& 857	& 946	& 841\\
$T_{97}$ & 1076	& 1120	& 1120	& 1156\\
\hline
\multicolumn{4}{l}{Two-parameter fits}\\
$T_w$ (K)                     & 382	& 558	& 470	&  422 	\\
log$_{10} \, (N_w$/cm$^{-2})$ & 19.89  	& 19.80	& 19.84	&  19.99\\
OPR$_w$                       & 0.43	& 0.41	& 0.48	&  0.21	\\
\\
$T_h$ (K)                       &1025	& 1150	& 1062	&  1029 	\\
log$_{10} \, (N_h$/cm$^{-2})$   &19.22	& 18.86	& 19.13	&  19.17	\\
OPR$_h$                         &1.53	& 1.55	& 1.98	&  1.88	\\
\hline
log$_{10} \, (N_w + N_h$)/cm$^{-2}$)  & 19.97	& 19.84	& 19.92	& 20.05  	\\
$N_{4-9}$ (cm$^{-2}$)                 &	18.83   & 18.88	& 18.91	& 18.93 	\\

\enddata
\end{deluxetable}

\clearpage

\begin{deluxetable}{lccccc}
\tablewidth{0pt}
\tablecaption{{\em Spitzer}-Derived H$_2$O Abundance Upper Limits}
\tablehead{&&\multicolumn{2}{c}{7$_{25}$ - 6$_{16}$ (29.837\um)} &
\multicolumn{2}{c}{6$_{34}$ - 5$_{05}$ (30.899\um)}\\\hline
Source & H$_2$ density (cm$^{-3}$) & Line intensity$^1$ & $n({\rm H_2O})/n({\rm H_2})$ &  Line intensity$^1$ 
& $n({\rm H_2O})/n({\rm H_2})$\\}
\startdata
\rule{0mm}{6mm}~HH54 FS~ & 10$^5$ & $<\:$9 & 
 \phantom{$^{\,\rm a}$}$<\:$1.6 $\times 10^{-4}$ & 
 $<\:$9 & $<\:$1.6 $\times 10^{-4}$ \\
 & 10$^4$ & $<\:$9 & $<\:$1.7 $\times 10^{-3}$ & 
 $<\:$9 & $<\:$1.5 $\times 10^{-3}$\\ \hline
\rule{0mm}{6mm}~HH54 C & 10$^5$ & $<\:$6 & $<\:$1.1 $\times 10^{-4}$ & 
 $<\:$4 & $<\:$6.6  $\times 10^{-5}$\\
 & 10$^4$ & $<\:$6 & $<\:$1.2 $\times 10^{-3}$ & 
 $<\:$4 & $<\:$6.1 $\times 10^{-4}$ \\ \hline
\rule{0mm}{6mm}~HH54 E+K & 10$^5$ & $<\:$7 & $<\:$1.2 $\times 10^{-4}$ & 
 $<\:$9 & $<\:$1.5 $\times 10^{-4}$ \\
 & 10$^4$ & $<\:$7 & $<\:$1.3 $\times 10^{-3}$ & 
 $<\:$9 & $<\:$1.4  $\times 10^{-3}$ \\ \hline
\rule{0mm}{6mm}~HH7 & 10$^5$ & $<\:$10 & $<\:$1.8 $\times 10^{-4}$ & 
 $<\:$14 & $<\:$2.5 $\times 10^{-4}$\\
 & 10$^4$ & $<\:$10 & $<\:$1.9  $\times 10^{-3}$& 
 $<\:$14 & $<\:$2.3  $\times 10^{-3}$\\*[0.6mm] 
\enddata
\tablenotetext{1}{3$\sigma$  upper limits in units of $\rm 10^{-7}\,erg\,cm^{-2}\,s^{-1}\,sr^{-1}$}
\end{deluxetable}
\clearpage

\begin{deluxetable}{ccc}
\tablewidth{0pt}
\tablecaption{Effective beam size and centroid accuracy}
\tablehead{
Wavelength & Centroid offset & Half-power beam width \\
$\mu$m & $^{\prime\prime}$ & $^{\prime\prime}$ \\
& (R.A., Dec.) & (circular average)}

\startdata
5.8 & ( +0.5, --1.9 ) & 3.0 \\
9.2 & ( --0.6, --2.3 ) & 3.1 \\
10.5 & ( +0.1, +1.6 ) & 5.0 \\
18.5 & ( --0.5, +2.0 ) & 5.3 \\
22.0 & ( --0.2, +0.0 ) & 9.6 \\
30.5 & ( --0.3, --0.4 ) & 10.2 \\
\enddata
\end{deluxetable}

\begin{figure}
\includegraphics[scale=0.9,angle=0]{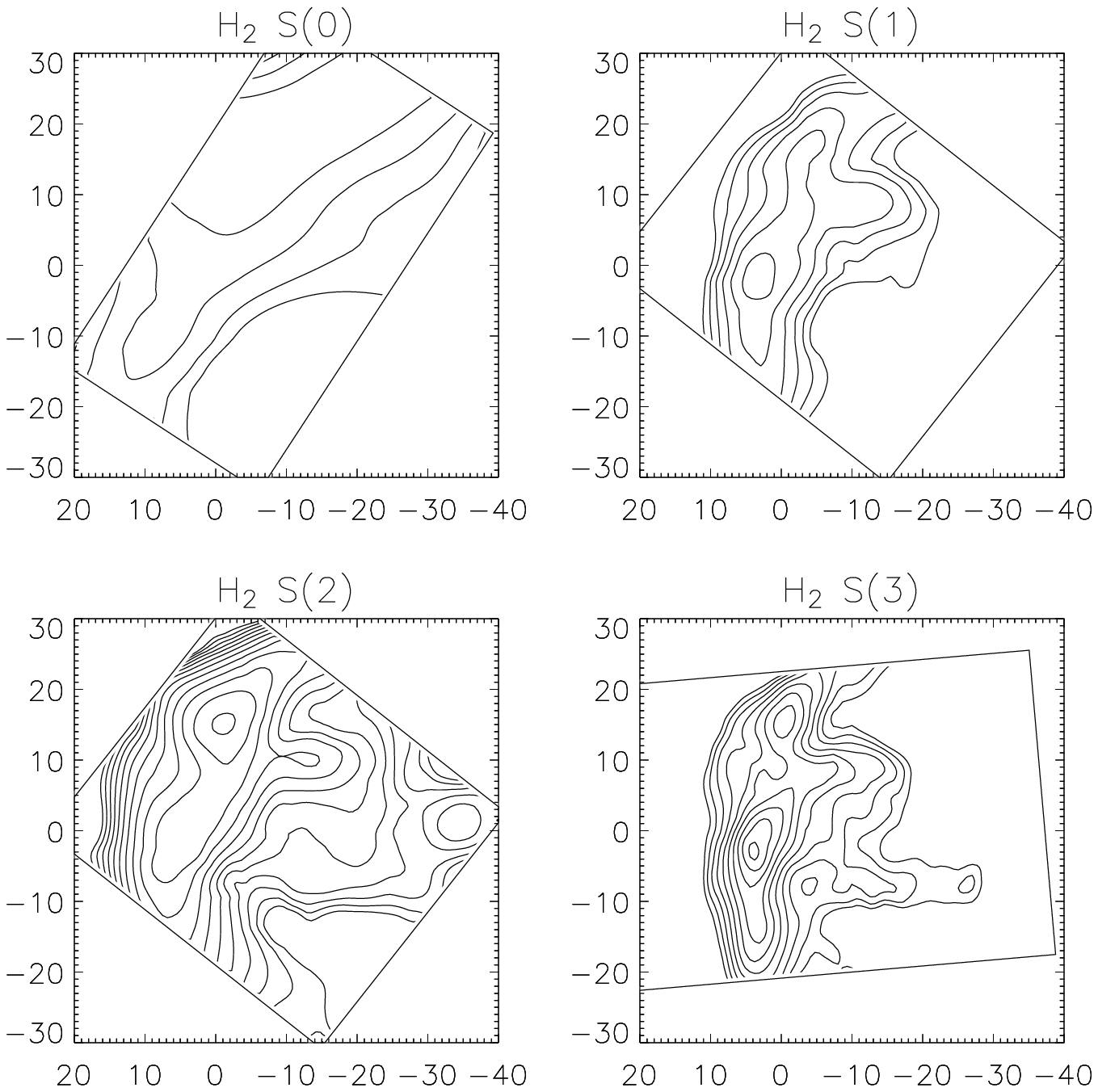}
\figcaption{H$_2$ S(0) -- S(3) emission line intensities observed toward HH54.  Logarithmic contours are shown, in steps of 0.1 dex, with the lowest contours corresponding to line intensities of $3 \times 10^{-6}$, 
$2 \times 10^{-5}$, $3 \times 10^{-5}$, and $1 \times 10^{-4} \rm \, erg \, \, cm^{-2} \, s^{-1} \, sr^{-1}$ 
respectively for H$_2$ S(0), S(1), S(2) and S(3).  The straight lines demark the regions within which each transition was mapped.  The horizontal and vertical axes show the R.A. ($\Delta \alpha \rm \cos \delta$) and declination ($\Delta \delta$) offsets in arcsec relative to $\alpha=$12h\,55m\,53.40s, $\delta=-76$d\,56$^\prime$20.5$^\prime$$^\prime$ (J2000).}

\end{figure}

\clearpage

\begin{figure}
\includegraphics[scale=0.9,angle=0]{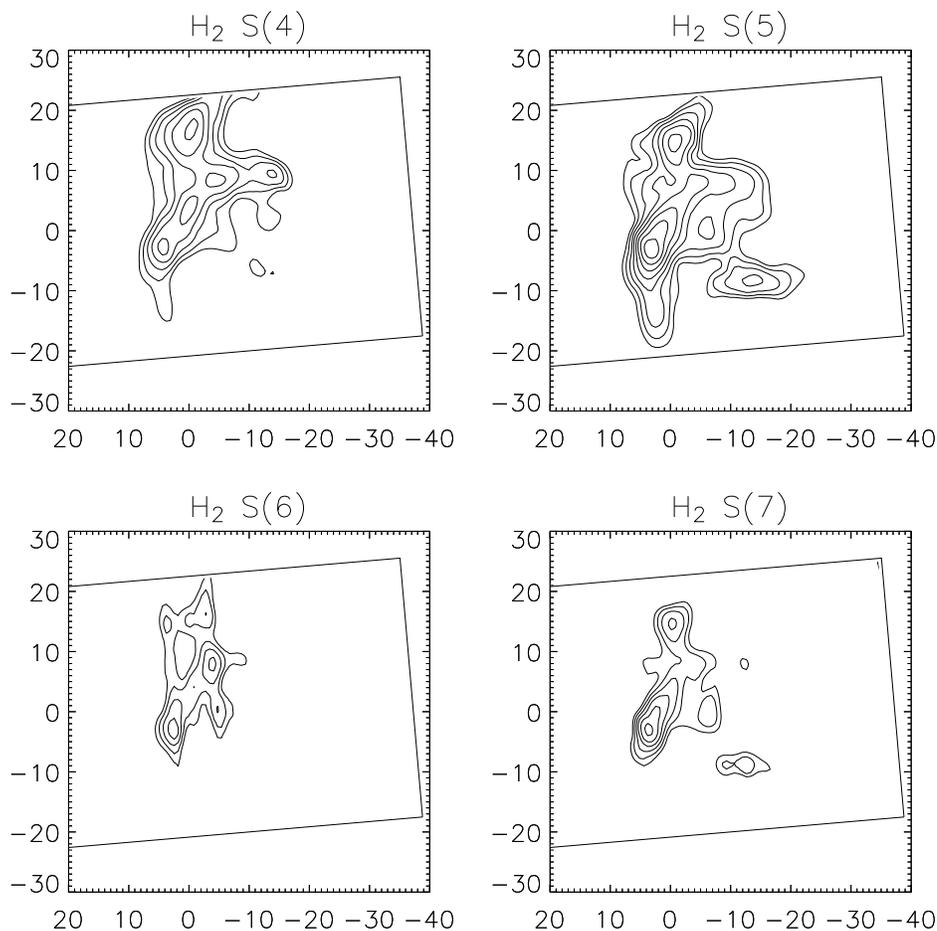}
\figcaption{H$_2$ S(4) -- S(7) emission line intensities observed toward HH54.  Logarithmic contours are shown, in steps of 0.1 dex, with the lowest contours corresponding to a line intensity of $2 \times 10^{-4} \rm \, \, erg \, cm^{-2} \, s^{-1} \, sr^{-1}$. The straight lines demark the regions within which each transition was mapped.  The horizontal and vertical axes show the R.A. ($\Delta \alpha \rm \cos \delta$) and declination ($\Delta \delta$) offsets in arcsec relative to $\alpha=$12h\,55m\,53.40s, $\delta=-76$d\,56$^\prime$20.5$^\prime$$^\prime$ (J2000).}
\end{figure}

\clearpage

\begin{figure}
\includegraphics[scale=0.9,angle=0]{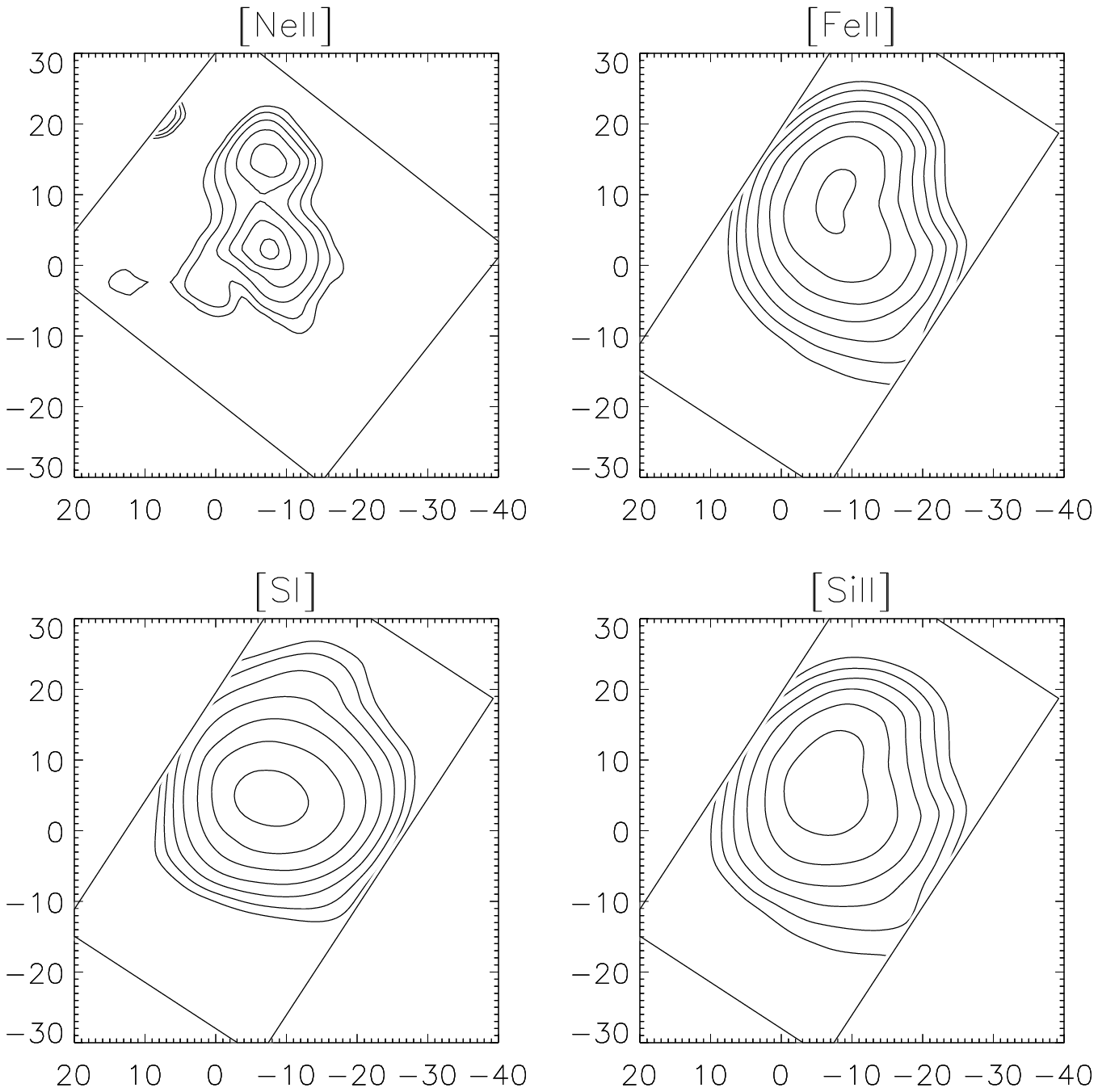}
\figcaption{Fine structure emission line intensities observed toward HH54.  Logarithmic contours are shown, in steps of 0.1 dex, with the lowest contours corresponding to line intensities of $1 \times 10^{-5}$, 
$1 \times 10^{-5}$, $5 \times 10^{-6}$, and $2 \times 10^{-5} \rm \, erg \, \, cm^{-2} \, s^{-1} \, sr^{-1}$ 
respectively for the [NeII] $12.8\,\mu$m, [FeII] $26\,\mu$m, [SI] $25\,\mu$m and [SiII] $35\,\mu$m transitions.  The straight lines demark the regions within which each transition was mapped.  The horizontal and vertical axes show the R.A. ($\Delta \alpha \rm \cos \delta$) and declination ($\Delta \delta$) offsets in arcsec relative to $\alpha=$12h\,55m\,53.40s, $\delta=-76$d\,56$^\prime$20.5$^\prime$$^\prime$ (J2000).}
\end{figure}

\clearpage

\begin{figure}
\includegraphics[scale=0.8,angle=0]{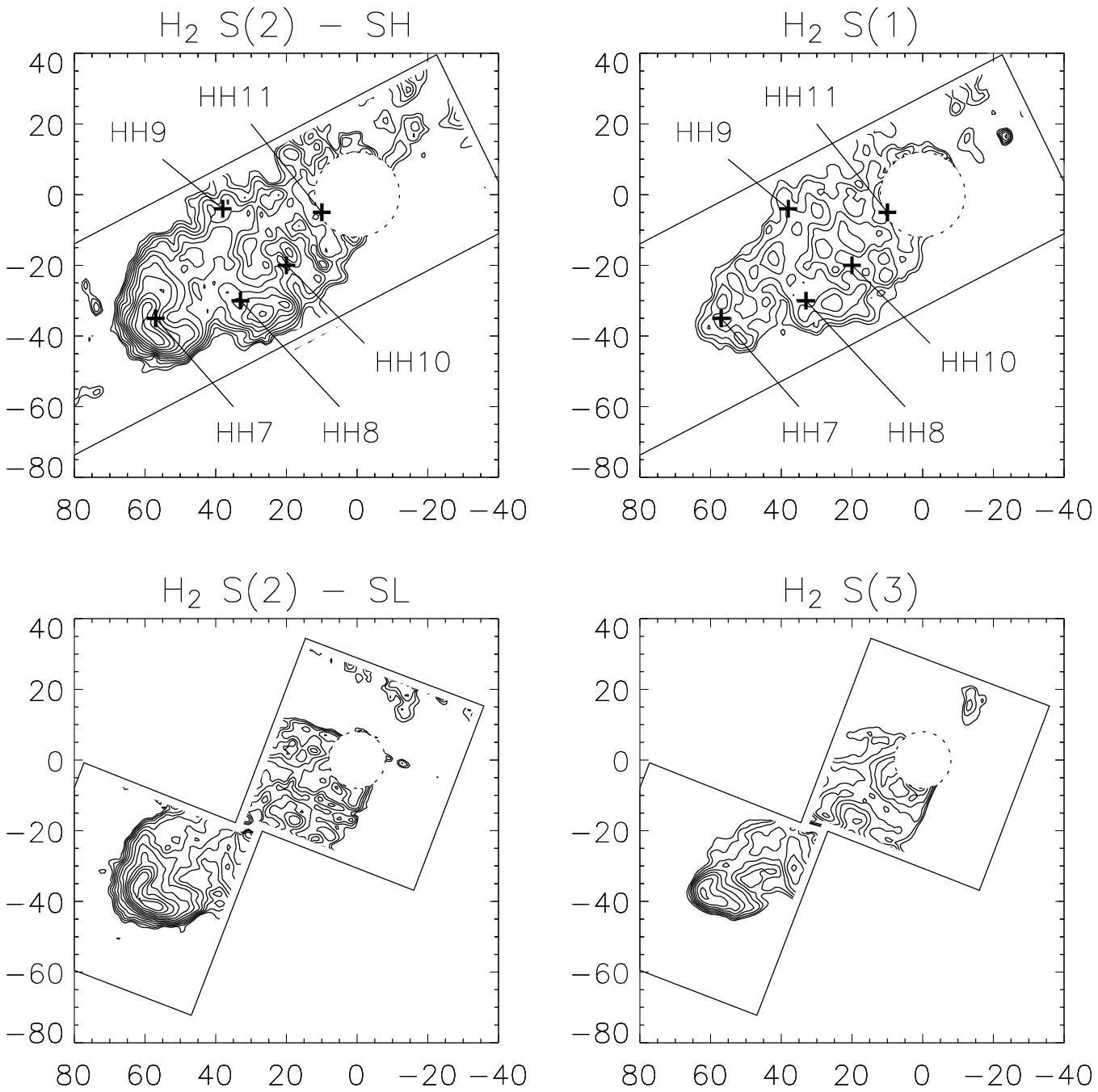}
\figcaption{H$_2$ S(1) -- S(3) emission line intensities observed toward HH7--11.  Logarithmic contours are shown, in steps of 0.1 dex, with the lowest contours corresponding to line intensities of $5 \times 10^{-6}$, 
$3 \times 10^{-5}$, $3 \times 10^{-5}$, and $1 \times 10^{-4} \rm \, \, erg \, cm^{-2} \, s^{-1} \, sr^{-1}$ 
respectively for H$_2$ S(1), S(2) and S(3).  The straight lines demark the regions within which each transition was mapped, and the dashed circles centered at offset (0,0) encircle the regions close to the bright continuum source SVS13 where the line intensity estimates are unreliable.  The upper map labeled H$_2$ S(2) was obtained with the Short-High module and the lower map with the Short-Low module.  The horizontal and vertical axes show the R.A. ($\Delta \alpha \rm \cos \delta$) and declination ($\Delta \delta$) offsets in arcsec relative to $\alpha=$3h\,29m\,3.74s, $\delta=+31$d\,16$^\prime$04.0$^\prime$$^\prime$ (J2000).  The marked positions in the upper two panels are based upon the observations of Khanzadyan et al.\ (2003).}

\end{figure}

\clearpage

\begin{figure}
\includegraphics[scale=0.9,angle=0]{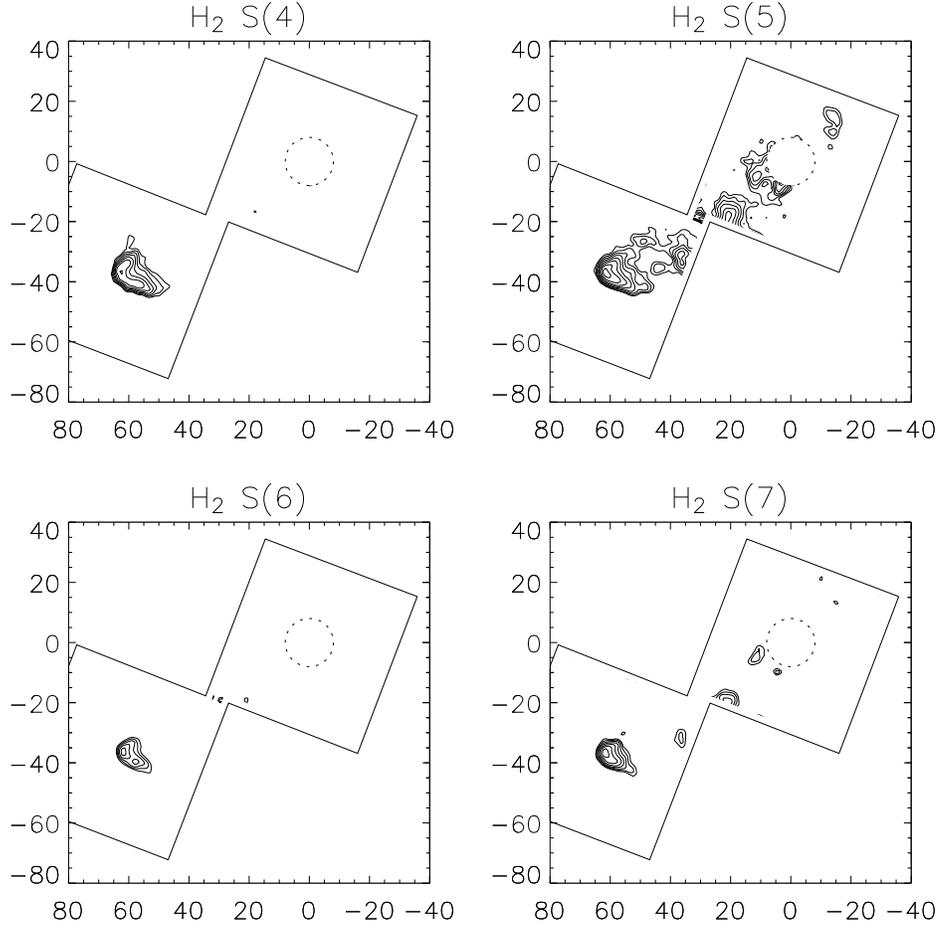}
\figcaption{H$_2$ S(4) -- S(7) emission line intensities observed toward HH7--11.  Logarithmic contours are shown, in steps of 0.1 dex, with the lowest contours corresponding to line intensities of $2 \times 10^{-4} \rm \, \, erg \, cm^{-2} \, s^{-1} \, sr^{-1}$.  The straight lines demark the regions within which each transition was mapped, and the dashed circles centered at offset (0,0) encircle the regions close to the bright continuum source SVS13 where the line intensity estimates are unreliable.  The horizontal and vertical axes show the R.A. ($\Delta \alpha \rm \cos \delta$) and declination ($\Delta \delta$) offsets in arcsec relative to $\alpha=$3h\,29m\,3.74s, $\delta=+31$d\,16$^\prime$04.0$^\prime$$^\prime$ (J2000).}

\end{figure}

\clearpage

\begin{figure}
\includegraphics[scale=0.9,angle=0]{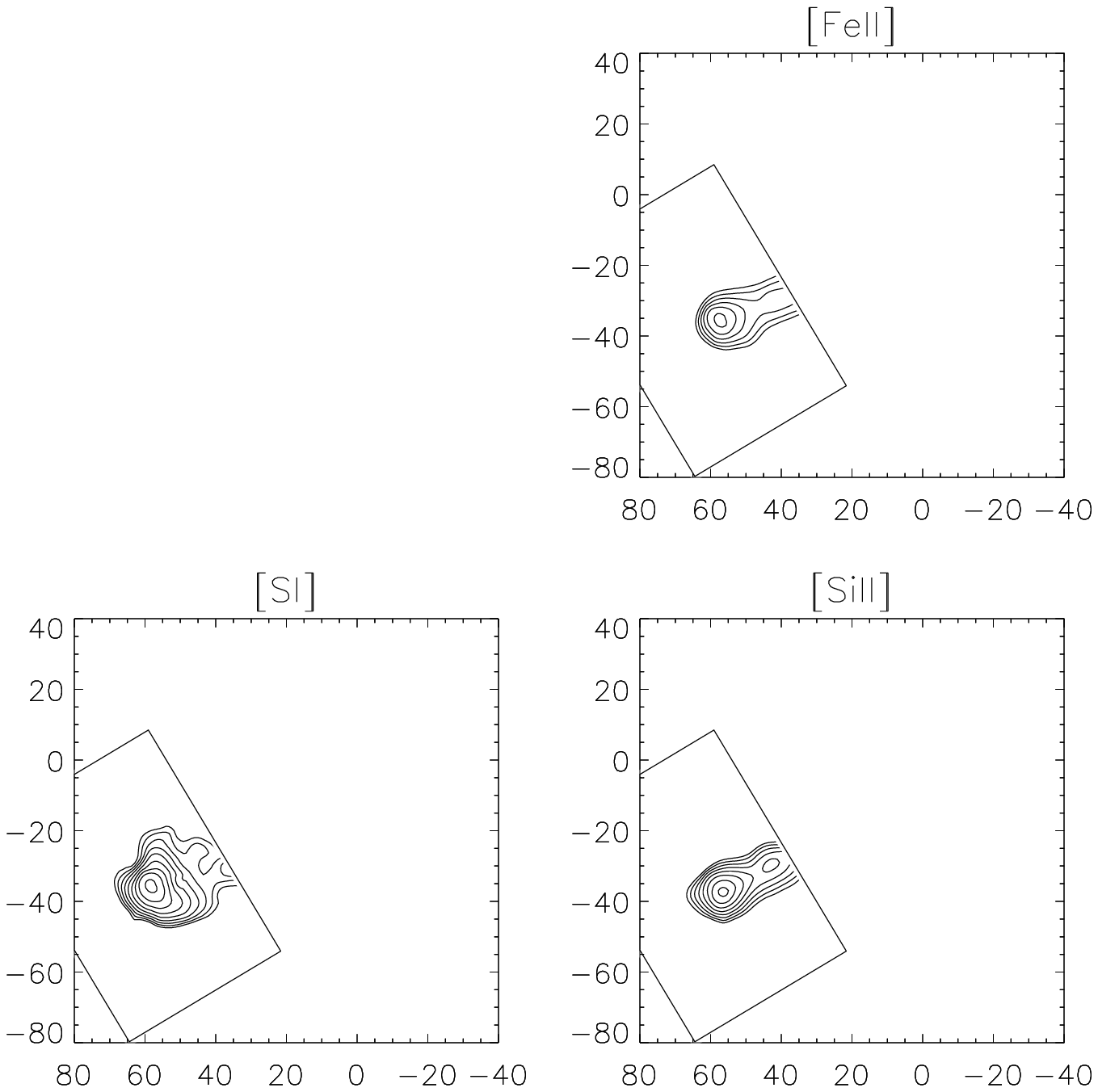}
\figcaption{Fine structure emission line intensities observed toward HH7--11.  Logarithmic contours are shown, in steps of 0.1 dex, with the lowest contours corresponding to line intensities of 
$1 \times 10^{-5}$, $5 \times 10^{-6}$, and $2 \times 10^{-5} \rm \, erg \, \, cm^{-2} \, s^{-1} \, sr^{-1}$ 
respectively for the [FeII] $26\,\mu$m, [SI] $25\,\mu$m and [SiII] $35\,\mu$m transitions.  The straight lines demark the regions within which each transition was mapped.  The horizontal and vertical axes show the R.A. ($\Delta \alpha \rm \cos \delta$) and declination ($\Delta \delta$) offsets in arcsec relative to $\alpha=$3h\,29m\,3.74s, $\delta=+31$d\,16$^\prime$04.0$^\prime$$^\prime$ (J2000).}
\end{figure}

\clearpage

\begin{figure}
\includegraphics[scale=1.0,angle=0]{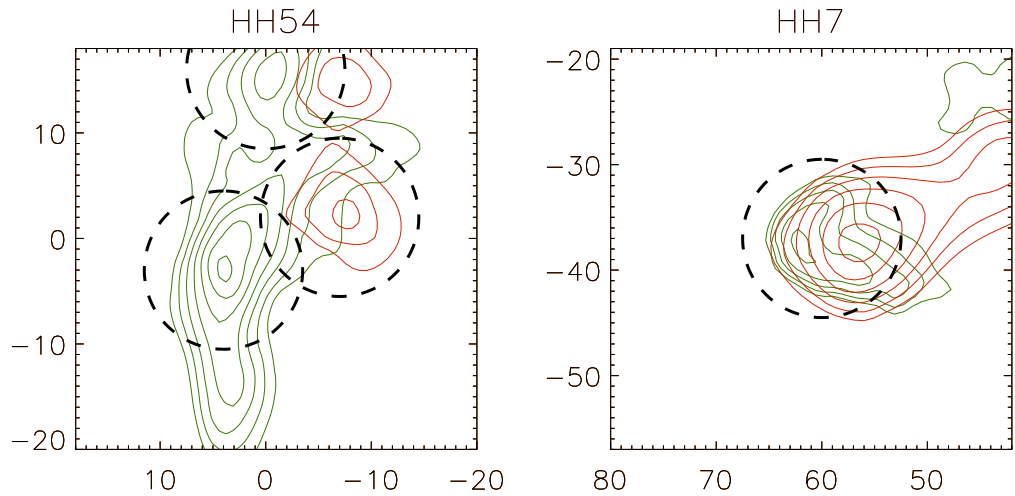}
\figcaption{Comparison between the distribution of H$_2$ and fine-structure emission lines.
Green contours:  H$_2$ S(3) line; red contours [NeII] $12.8\,\mu$m line (HH54) or [SiII] $35\,\mu$m line (HH7)
Logarithmic contours are shown, in steps of 0.1 dex, with the lowest contours corresponding to line intensities of 
$2 \times 10^{-4}$, $2 \times 10^{-5}$ and $3 \times 10^{-5} \rm \, erg \, \, cm^{-2} \, s^{-1} \, sr^{-1}$
respectively for the H$_2$ S(3), [NeII] and [SiII] transitions.  The dashed circles indicate the apertures for which average spectra are shown in Figure 8: these are denoted HH54E+K, HH54FS, HH54C (left panel from south to north) and HH7 (right panel).}
\end{figure}

\clearpage

\begin{figure}
\includegraphics[scale=0.65,angle=270]{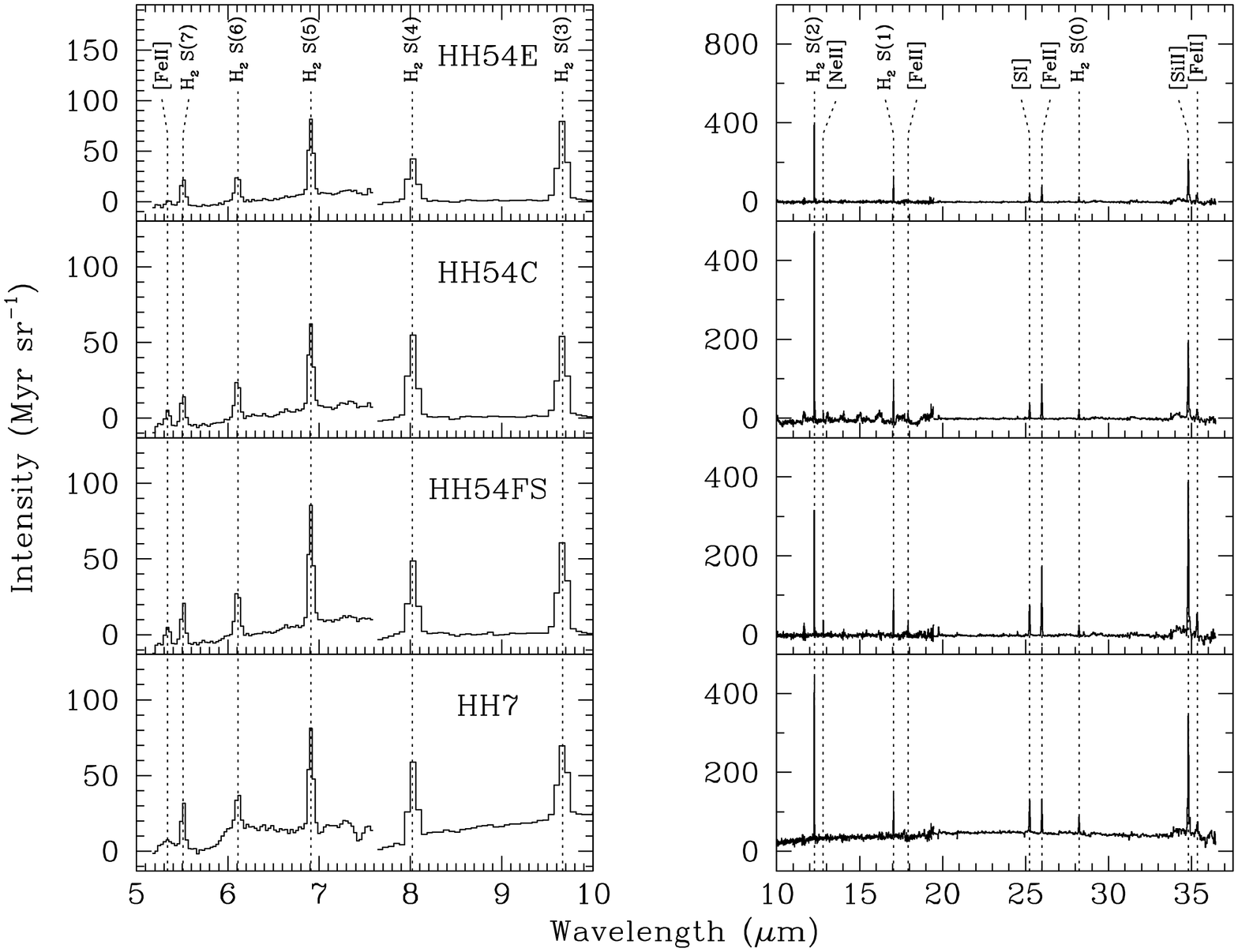}
\figcaption{Average spectra observed for $15^{\prime\prime}$ (HPBW) diameter circular apertures centered at offsets $(\Delta \alpha \rm \cos \delta, \Delta \delta) = (-7^{\prime\prime},+2^{\prime\prime})$ (HH54FS), $(0^{\prime\prime},+16^{\prime\prime})$ (HH54C) and $(4^{\prime\prime},-3^{\prime\prime})$ (HH54E+K) relative to $\alpha=$12h\,55m\,53.40s, $\delta=-76$d\,56$^\prime$20.5$^\prime$$^\prime$ (J2000), and offset $(\Delta \alpha \rm \cos \delta, \Delta \delta) = (+60^{\prime\prime},-37^{\prime\prime})$ (HH7) relative to $\alpha=$3h\,29m\,3.74s, $\delta=+31$d\,16$^\prime$04.0$^\prime$$^\prime$ (J2000).  The spectra in the left panel were obtained with the SL module, while those in the right panel were obtained with SH ($\lambda \le \rm 19.5\,\mu m$) and LH ($\lambda \ge \rm 19.5\,\mu m$).}
\end{figure}

\clearpage

\begin{figure}
\includegraphics[scale=0.6,angle=0]{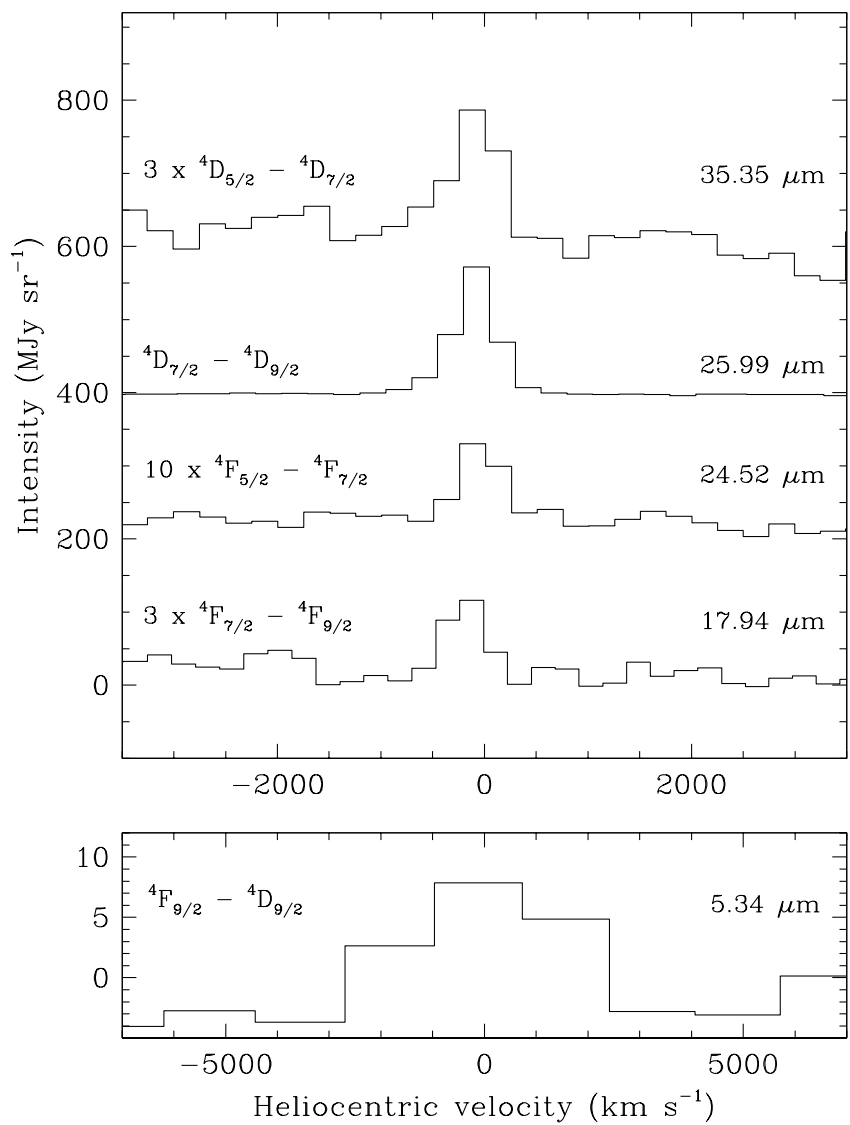}
\figcaption{[FeII] fine-structure lines detected toward HH54.  The spectra are averages for a $15^{\prime\prime}$ (HPBW) diameter circular aperture centered at offset $(\Delta \alpha \rm \cos \delta, \Delta \delta) = (-7^{\prime\prime},2^{\prime\prime})$ (position HH54FS) relative to $\alpha=$12h\,55m\,53.40s, $\delta=-76$d\,56$^\prime$20.5$^\prime$$^\prime$ (J2000). Arbitrary continuum offsets have been introduced for clarity.}
\end{figure}

\clearpage

\begin{figure}
\includegraphics[scale=0.8,angle=0]{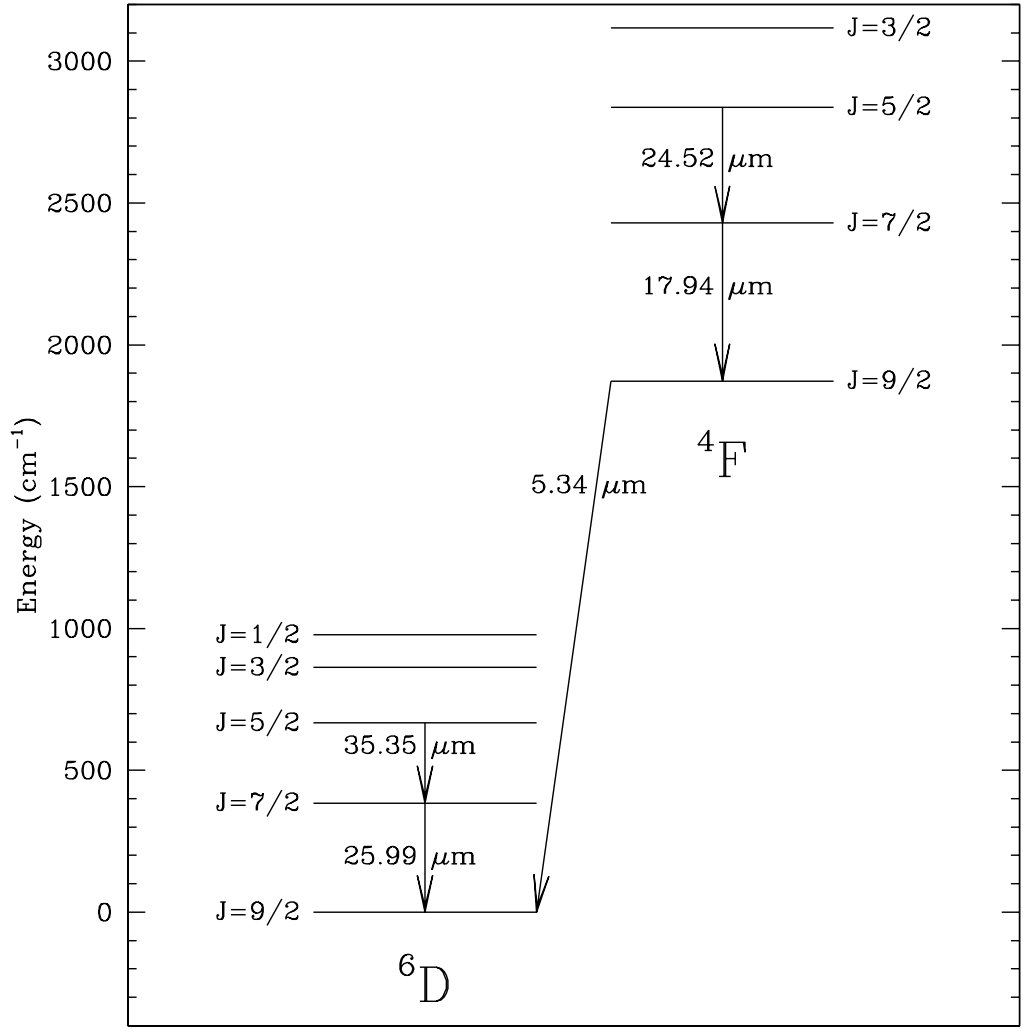}
\figcaption{Grotrian diagram for [FeII], showing the five fine-structure transitions detected toward HH54}
\end{figure}

\clearpage

\begin{figure}
\includegraphics[scale=0.6,angle=0]{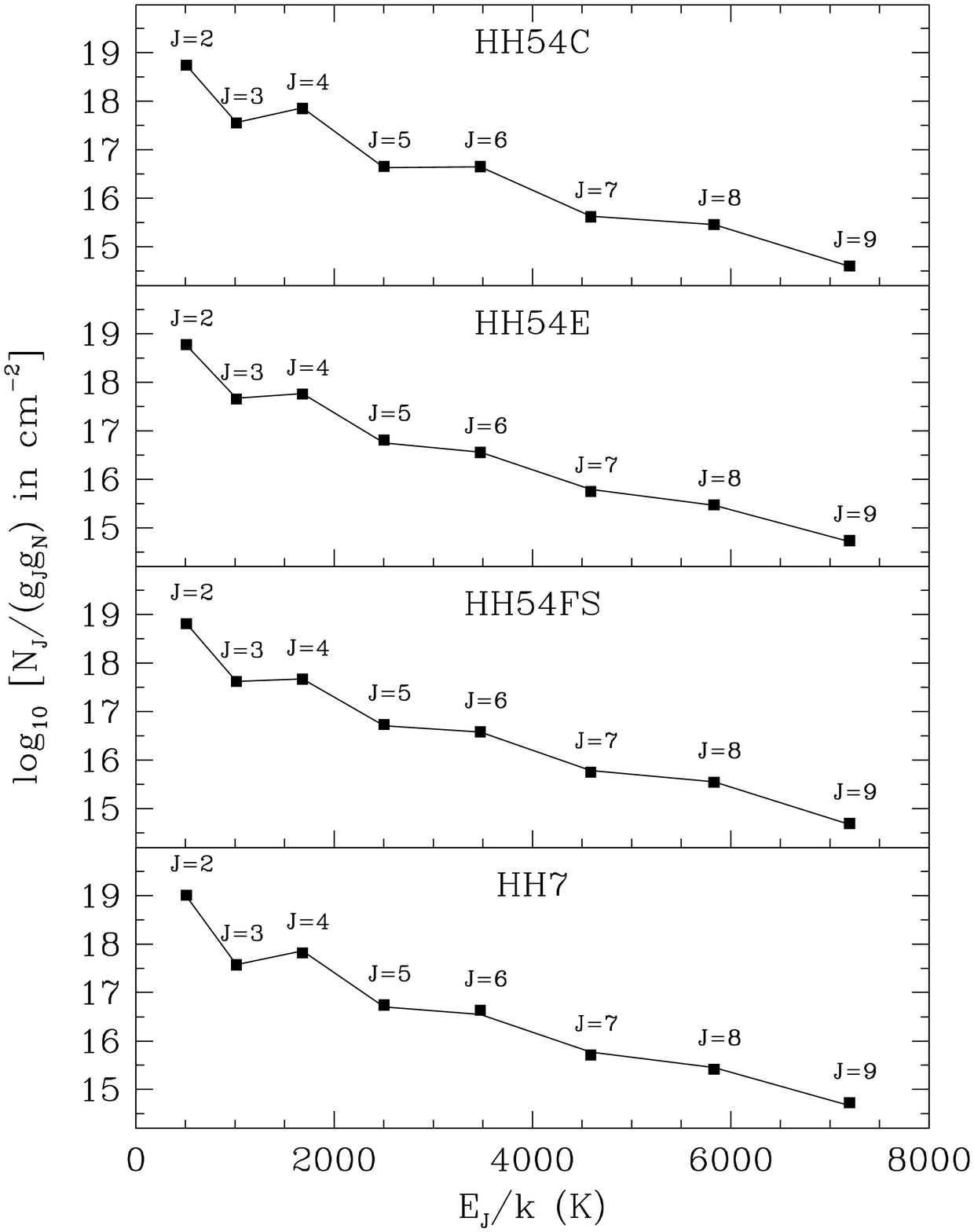}
\figcaption{H$_2$ rotational diagrams, computed for $15^{\prime\prime}$ (HPBW) diameter circular apertures centered at offsets $(\Delta \alpha \rm \cos \delta, \Delta \delta) = (-7^{\prime\prime},+2^{\prime\prime})$ (HH54FS), $(0^{\prime\prime},+16^{\prime\prime})$ (HH54C) and $(4^{\prime\prime},-3^{\prime\prime})$ (HH54E+K) relative to $\alpha=$12h\,55m\,53.40s, $\delta=-76$d\,56$^\prime$20.5$^\prime$$^\prime$ (J2000), and offset $(\Delta \alpha \rm \cos \delta, \Delta \delta) = (+60^{\prime\prime},-37^{\prime\prime})$ (HH7) relative to $\alpha=$3h\,29m\,3.74s, $\delta=+31$d\,16$^\prime$04.0$^\prime$$^\prime$ (J2000).  Square symbols indicate the measured values, while the lines show the best two-component fit to the observations.}
\end{figure}

\clearpage

\begin{figure}
\includegraphics[scale=0.6,angle=0]{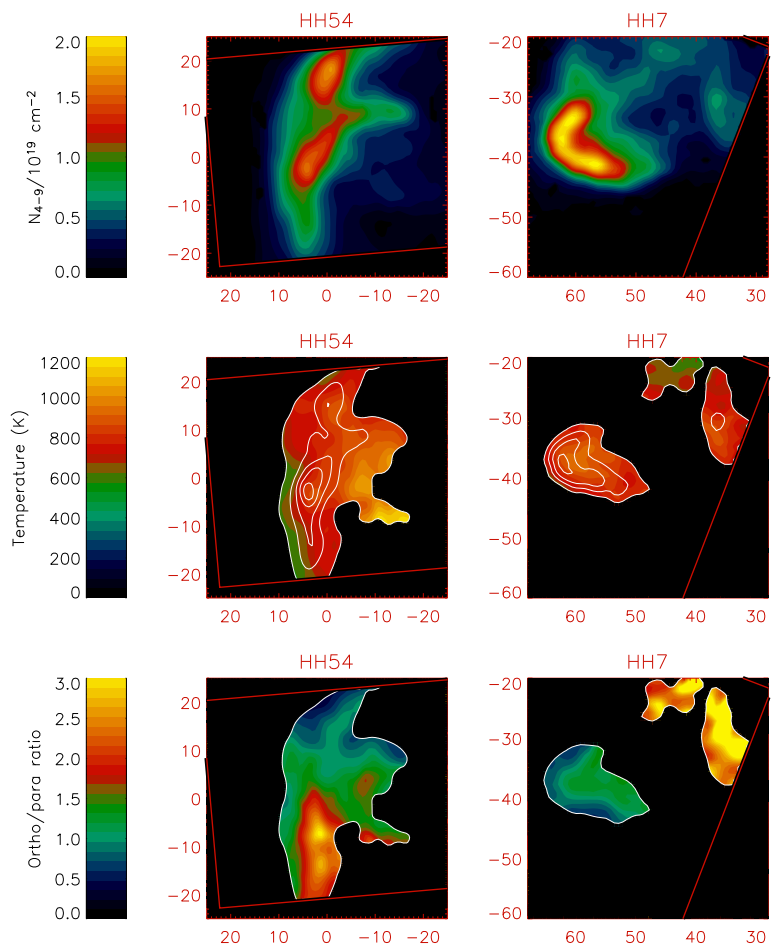}
\figcaption{Maps of the H$_2$ column density (top panels), H$_2$ rotational temperature (middle) and H$_2$ ortho-to-para ratio (bottom panels) derived from a least-squares fit to the $J=4$, 5, 6, and 7 column densities.  Superposed white contours (middle panels) show the H$_2$ S(3) line intensity (contours at 3.0, 4.5, 6.0, 7.5 $\times 10^{-4}\rm \, erg \, \, cm^{-2} \, s^{-1} \, sr^{-1}$ for HH54 and  2.5, 3.5, 4.5, 5.5 $\times 10^{-4}\rm \, erg \, \, cm^{-2} \, s^{-1} \, sr^{-1}$ for HH7.)  Regions where the signal-to-noise ratio is inadequate to yield a fit to the H$_2$ rotational diagram appear in black.}  
\end{figure}

\clearpage

\begin{figure}
\includegraphics[scale=0.6,angle=0]{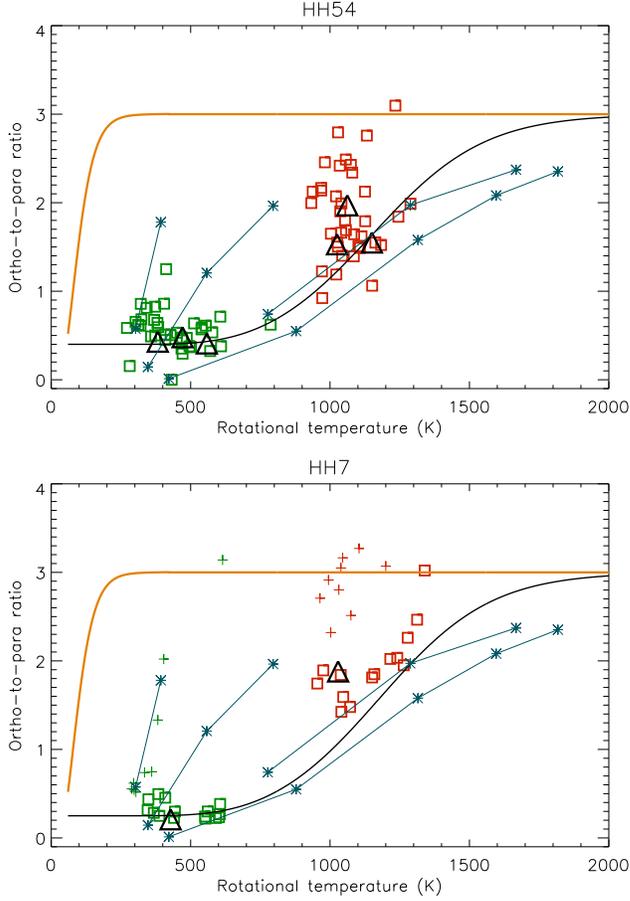}
\figcaption{Correlation between ortho-to-para ratio and rotational temperature, obtained from two-component fits 
to the $J=2$, 3, 4, 5, 6, 7, 8 and 9 column densities.  Triangles are from a two-component fit to the rotational diagrams shown in Figure 11.  Other data points apply to $5 \times 5$ arcsec square subregions within the source regions shown in Figure 12, green and red referring respectively to the warm and hot gas components. For HH7, 
the squares apply to the HH7 bow shock, while crosses apply to warm gas $\sim 20$ arcsec behind the bow shock (HH8 and nearby warm gas).  The black curve shows the expected behavior for gas with an initial ortho-to-para ratio of 0.25 (lower panel) or 0.4 (upper panel) that has been warm for a period of $150/n_2({\rm H})$ yr, where $100\,n_2({\rm H})$ is the density of atomic hydrogen.  The orange curve shows the ortho-to-para ratio in LTE.  The cyan curves and asterisks show the model predictions of Wilgenbus et al.\ (2000; see text).} 
\end{figure}

\clearpage

\begin{figure}
\includegraphics[scale=0.9,angle=0]{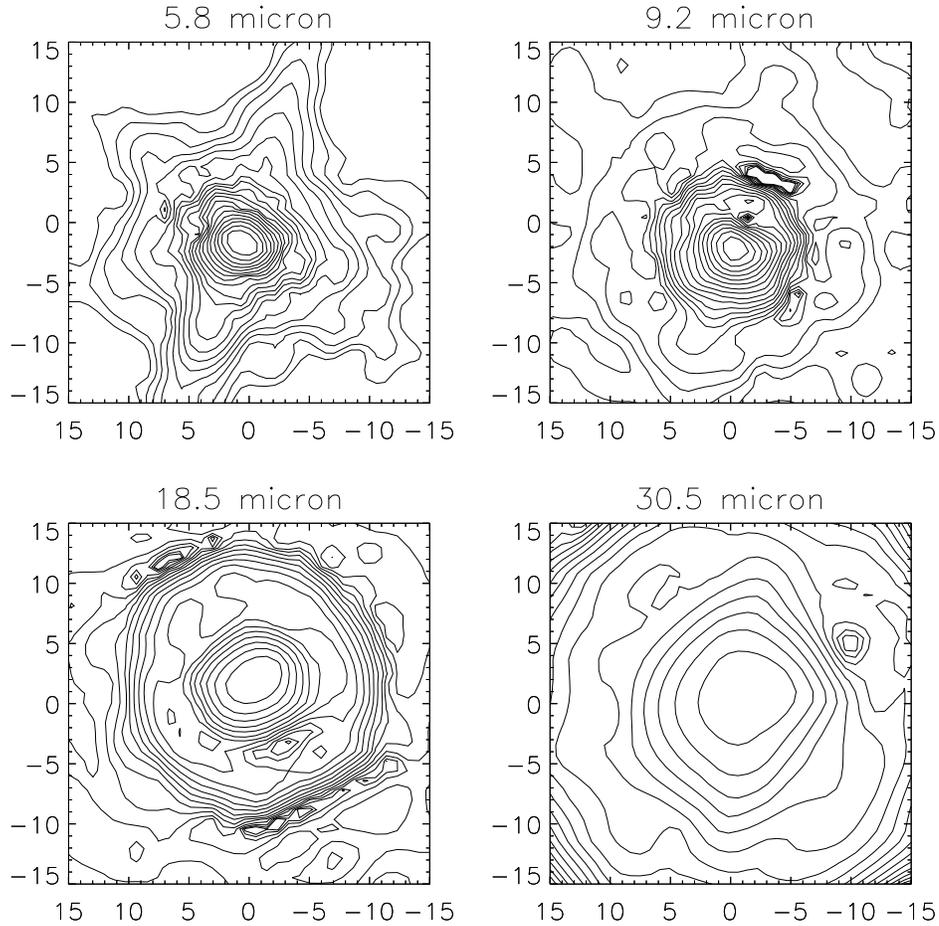}
\figcaption{Continuum maps obtained for several continuum wavelengths toward HH7--11 Field 1, which contains the strong continuum source SVS13 at its center (i.e. at position (0,0)).  Logarithmic contours are spaced by a factor $2^{1/2}$, with the second-highest contour corresponding to an intensity one-half the maximum.  To the extent that SVS13 can be regarded as an unresolved point source at these wavelengths, these maps represent the beam profile resulting at the end of the entire data reduction procedure.} 
\end{figure}

\clearpage

\begin{figure}
\includegraphics[scale=0.6,angle=270]{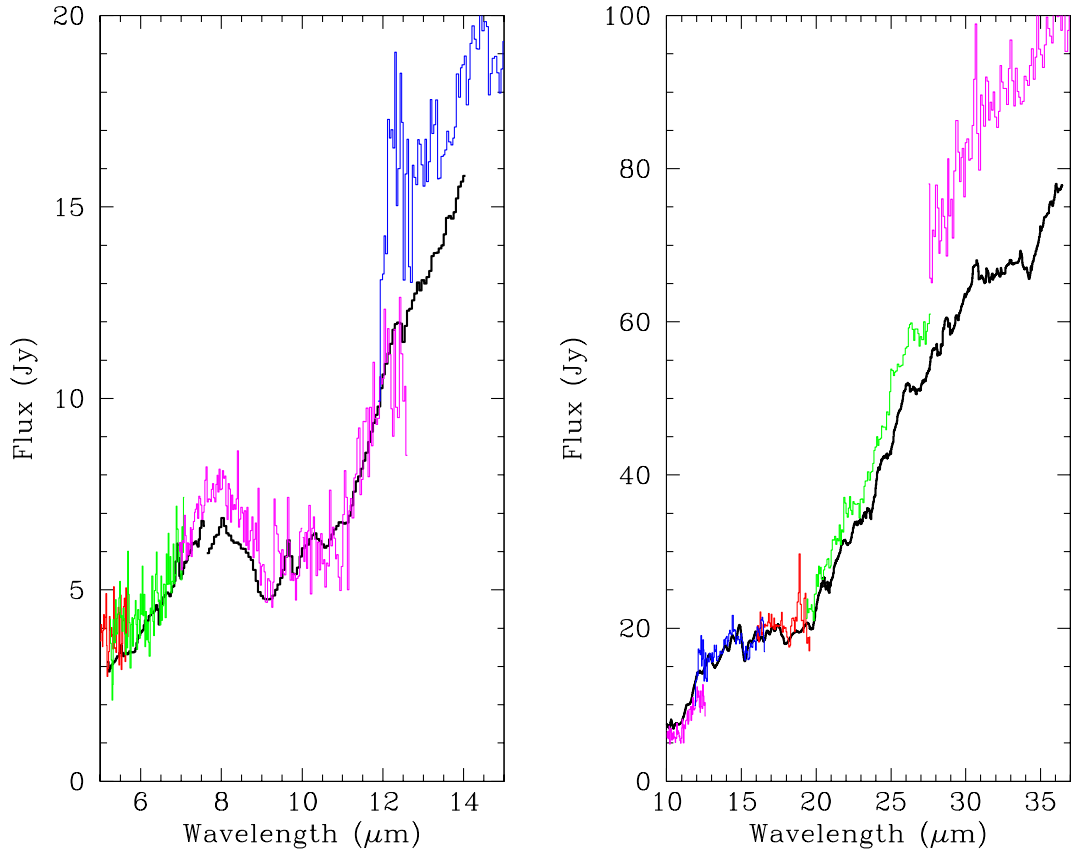}
\figcaption{Comparison of the SVS13 spectrum obtained with {\it Spitzer} (heavy black line) with that obtained with {\it ISO} (red, blue, green and magenta lines, different colors being used to show the different SWS orders).  The {\it Spitzer} spectrum obtained with the Short-Low module appears in the left panel, and that obtained with the Short-High and Long-High modules appears in the right panel.}
\end{figure}

\clearpage

\begin{figure}
\includegraphics[scale=0.75,angle=0]{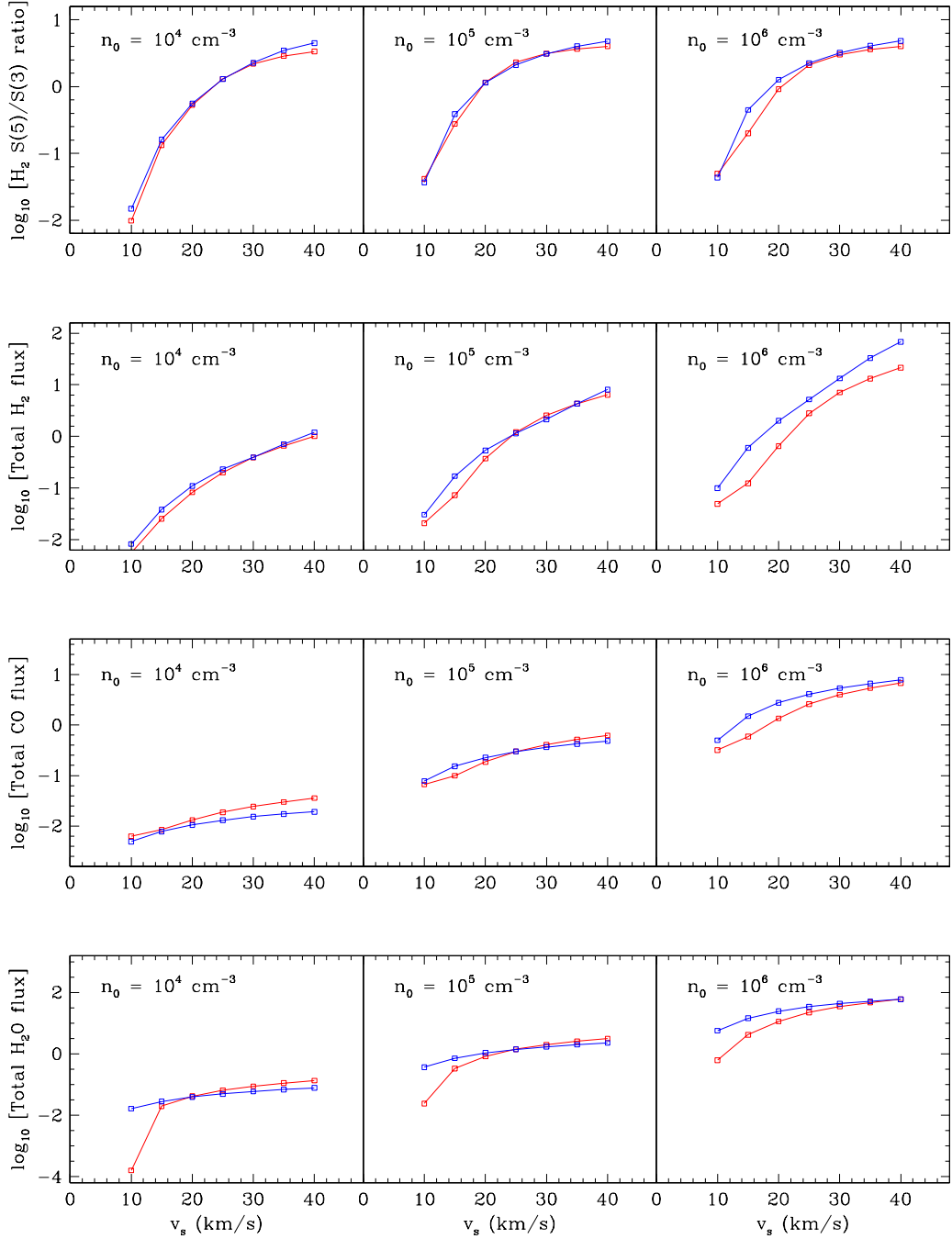}
\figcaption{Comparison of the H$_2$ S(5)/S(3) line ratio and the total H$_2$, H$_2$O and CO surface luminosities (in erg~cm$^{-2}$~s$^{-1}$) derived for the simple slab model (red; see text) with those obtained in the detailed calculations of KN96 (blue).}  
\end{figure}

\end{document}